\documentclass[12pt]{article}
\usepackage[a4paper]{geometry}
\usepackage{babel}

\usepackage{fancyhdr}
\usepackage{amsmath}
\usepackage{amsfonts}
\usepackage{amstext}
\usepackage{amssymb}
\usepackage{amsthm}
\usepackage{amscd}

\usepackage[pagebackref,draft=false]{hyperref}
\hypersetup{colorlinks,
linkcolor=myrefcolor,
citecolor=mycitecolor,
urlcolor=myurlcolor}

\usepackage[capitalize]{cleveref}
\usepackage{caption}
\usepackage{etaremune}

\usepackage{xcolor}
\definecolor{myurlcolor}{rgb}{0,0,0.4}
\definecolor{mycitecolor}{rgb}{0,0.5,0}
\definecolor{myrefcolor}{rgb}{0.5,0,0}
\usepackage{graphicx}
\usepackage{tikz}
\usepackage{tikz-cd}
\usepackage{mathrsfs}

\usepackage{etoolbox}
\usepackage{makeidx}
\usepackage{sectsty}
\usepackage{dsfont}
\usepackage{enumitem} 
\usepackage[]{latexsym}
\usepackage{braket}
\usepackage{caption}
\usepackage[utf8]{inputenx}
\usepackage[T1]{fontenc}
\usepackage{lmodern}
\usepackage{textcomp}
\usepackage{microtype}
\usepackage{totcount}
\usepackage{blindtext}

\newtheorem{remark}{Remark}

\newtheorem{corollary}{Corollary}
\newtheorem{proposition}{Proposition}
\newtheorem{definition}{Definition}

\newtheorem*{proof*}{Proof}

\newcommand{\be}{\begin{equation}}
\newcommand{\ee}{\end{equation}}
\newcommand{\bea}{\begin{eqnarray}}
\newcommand{\eea}{\end{eqnarray}}
\newcommand{\vsp}{\vspace{0.4cm}}
\newcommand{\grit}[1]{{\bfseries {\itshape {#1}}}}

\newcommand{\blue}[1]{\color{blue}{{#1}}}

\newcommand{\ra}{\rightarrow}

\newcommand{\cfr}[1]{({\itshape cf.} {#1})}

\newcommand{\hh}{\mathcal{H}}
\newcommand{\bh}{\mathcal{B}(\mathcal{H})}
\newcommand{\bhsa}{\mathcal{B}_{sa}(\mathcal{H})}
\newcommand{\bhpos}{\mathcal{B}_{+}(\mathcal{H})}

\newcommand{\Uh}{\mathcal{U}(\mathcal{H})}

\newcommand{\Tr}{\textit{Tr}}

\newcommand{\stsp}{\mathscr{S}}

\newcommand{\appa}{\mathscr{A}}
\newcommand{\appas}{\mathscr{A}_{sa}}

\newcommand{\bappa}{\mathscr{B}}

\newcommand{\pos}{\mathscr{P}}

\newcommand{\nplf}{{\itshape n.p.l.f.}}
\newcommand{\nplfs}{{\itshape n.p.l.f.s}}

\newcommand{\calg}{$C^{\star}$-algebra}
\newcommand{\calgs}{$C^{\star}$-algebras}
\newcommand{\walg}{$W^{\star}$-algebra}
\newcommand{\walgs}{$W^{\star}$-algebras}

\newcommand{\jj}{\mathrm{j}}
\newcommand{\gr}{\mathrm{g}}

\title{Parametric models and information geometry on W*-algebras}

\author{F. M. Ciaglia$^{1,4}$  \href{https://orcid.org/0000-0002-8987-1181}{\includegraphics[scale=0.7]{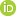}}, F. Di Nocera$^{2,5}$  \href{https://orcid.org/0000-0002-1415-2422}{\includegraphics[scale=0.7]{ORCID.png}} J. Jost$^{2,6}$\href{https://orcid.org/0000-0001-5258-6590}{\includegraphics[scale=0.7]{ORCID.png}}, L. Schwachh\"{o}fer$^{3,7}$\href{https://orcid.org/0000-0002-4268-6923}{\includegraphics[scale=0.7]{ORCID.png}}\\
\footnotesize{$^{1}$\textit{Departamento de Matem\'aticas, Universidad Carlos III de Madrid,}}  \\ \footnotesize{\textit{ Avenida de la Universidad 30, 28911, Legan\'es, Madrid, Spain.}} \\
\footnotesize{$^{2}$\textit{Max Planck Institut  für Mathematik in den Naturwissenschaften,}} \\   \footnotesize{\textit{ Inselstraße 22, 04103  Leipzig, Germany.}} \\
\footnotesize{$^{3}$\textit{Fakultät für Mathematik, Technische Universität Dortmund,}} \\   \footnotesize{\textit{  Vogelpothsweg 87,
44227, Dortmund, Germany.}} \\
\footnotesize{$^{4}$\textit{ e-mail: \texttt{f.ciaglia[at]math.uc3m.es}} $^{5}$\textit{ e-mail: \texttt{dinocer[at]mis.mpg.de}}} \\
\footnotesize{$^{6}$\textit{ e-mail: \texttt{jjost[at]mis.mpg.de}} $^{7}$\textit{ e-mail: \texttt{lschwach[at]math.tu-dortund.de  }}}
}

\begin{document}

\maketitle

\begin{abstract}
We introduce the notion of smooth parametric model  of normal positive linear functionals on possibly infinite-dimensional \walgs\ generalizing the notions of parametric models used in classical and quantum information geometry.
We then use the Jordan product naturally available in this context in order to define a Riemannian metric tensor on parametric models satsfying suitable regularity conditions.
This Riemannian metric tensor reduces to the Fisher-Rao metric tensor, or to the Fubini-Study metric tensor, or to the Bures-Helstrom metric tensor when suitable choices for the \walg\  and the models are made.
\end{abstract}

\tableofcontents
\thispagestyle{fancy}

\section{Introduction} \label{sec:introduction}

In classical and quantum information geometry one usually deals with parametrized subsets of probability distributions or of quantum states, respectively, colloquially referred to as {\itshape parametric models}.
A typical example in the classical context is given by the family of Gaussian probability distributions, or, in the quantum context, by the family of quantum coherent states.

From both the conceptual and practical point of view, there could be physical theoretical constraints leading to situations in which only a certain family of probability distributions or quantum states  can be modelled or physically realized (think again to Gaussian probability distributions and quantum coherent states), thus justifying  the choice to work with parametric models. 

From a purely mathematical point of view, on the other hand, the choice of working with parametric models is mandatory if we want to exploit the mathematical formalism of standard differential geometry \cite{A-M-R-1988,Jost-2017,Lang-1999}.
Indeed, both the space of probability distributions on a measurable outcome space and the space of quantum states identified with the space of density operators on a complex, separable Hilbert space do not posses the structure of smooth manifold.
Quite interestingly, this already happens in finite dimensions:
in the classical case, the space of probability distributions on a discrete and finite outcome space $\mathcal{X}_{n}$ (with $n$ elements)  can be naturally identified with the unit simplex in $\mathbb{R}^{n}$ which is a typical example of smooth manifold with corners \cite{Michor-1980}; in the quantum case, the space of quantum states  identified with the space of density operators on a finite-dimensional complex Hilbert space $\hh$ is a smooth manifold with boundary known as the Bloch ball when $\mathrm{dim}(\hh)=2$  \cite{B-Z-2006,G-K-M-2006}  and a stratified manifold when $\mathrm{dim}(\hh)>2$ \cite{DA-F-2021}.
In infinite dimensions, the situation is even worse given the technicalities associated with infinite-dimensional differential geometry.

While it can be argued that there are  approaches aiming to build an infinite-dimensional non-parametric theory both in the classical \cite{P-S-1995} and the quantum case \cite{Jencova-2006}, we believe that they truly are parametric models in which the parameters lie in an infinite-dimensional manifold.
Indeed, the seminal work by Pistone and Sempi \cite{P-S-1995} deals with the Banach manifold structure not on the whole space of probability distributions on a measure space, but rather on the space of all those probability distributions mutually absolutely continuous with respect to a given reference probability measure $\mu$.
This choice clearly selects what can be reasonably called a parametric model of probability distributions.
Something similar happens in the work by Jencova  \cite{Jencova-2006} in which a Banach manifold structure is given not to the whole space of states on a $W^{\star}$-algebra $\appa$, but rather to the space of faithful normal states on $\appa$.

Consequently, in order to use the tools of standard differential geometry, as it is customary in classical and quantum information geometry \cite{Amari-2016,A-N-2000,L-Y-L-W-2020,Paris-2009,Suzuki-2019}, we must necessarily accept the need to work with parametric models.
The classical case has been thoroughly and sistematically investigated also in the infinite-dimensional setting \cite{A-J-L-S-2015,A-J-L-S-2017,A-J-L-S-2018},  while, to the best of our knowledge, the information geometry of parametric models of  quantum states (especially in the infinite-dimensional setting) is still vastly unexplored.

The aim of this work is to start the exploration of this land, and to do it in such a way that both the classical and quantum case can be simultaneously handled.
The key to achieve such a unification is found in the theory of $W^{\star}$-algebras.
Indeed, as it will be thoroughly explained in the following, both probability distributions and quantum states can be  thought as states (in the functional analytic sense) on a suitable \walg\ $\appa$ which is Abelian in the classical case and non-Abelian in the quantum one.
It is worth mentioning that the use of \walgs\ to deal with probability distributions on a discrete and finite outcome space in the context of classical information geometry has been recently proposed \cite{G-R-2006,G-R-2006-02,G-N-2020}.
Moreover, the use of \walgs\ as a common arena for classical and quantum information geometry was also recently introduced by some of us in the finite-dimensional case \cite{C-J-S-2020-02,C-J-S-2020}.

In this work we aim to discuss in detail a definition of possibly infinite-dimensional parametric models in the context of \walgs\ which is well-adapted to classical and quantum information geometry.
Moreover, we will also explore the Riemannian aspects of  these models showing how well-known geometric structures like the Fisher-Rao metric tensor or the Bures-Helstrom metric tensor are actually connected with   algebraic operations in the \walg\ under consideration, in particular, with the Jordan product on its self-adjoint part.

It is important to note that, following the spirit of \cite{A-J-L-S-2017}, we introduce a notion of parametric model that does not require the normalization condition usually implemented when dealing with probability distributions and quantum states (i.e., the normalization of the total volume for probability distributions, or the normalization of the trace for quantum states).
This choice reflects the idea of considering the normalization condition a kind of {\itshape ad hoc} constraint which should really bear  no sensible information (for instance, the choice of 1 instead of 73 as a normalization constant is completely irrelevant).

The work is structured as follows.
In section \ref{sec: walgs and nplfs} we introduce all the background material we need on \walgs\ and their relation with classical and quantum information geometry.
In section \ref{sec: parametric models} we introduce the notion of smooth parametric model of normal positive linear functionals on a \walg\ $\appa$. 
This notion allows to treat essentially all the parametric models  used in classical and quantum information geometry in the same mathematical framework.
In section \ref{sec: jordan parametric models} we further embellish the notion of parametric model in order to deal with geometric aspects related with Riemannian geometry.
Essentially, we characterize all those parametric models for which the algebraic structure of $\appa$ leads to the definition of a Riemannian metric tensor suitably generalizing the Fisher-Rao metric tensor.
In section \ref{sec: examples} we discuss three meaningful examples.
First of all, we discuss the finite-dimensional case in which almost all the technical difficulties disappear and there is a clear link with the Jordan-algebra-analogue of Kirillov's theory of co-adjoint orbits.
Then, we discuss the case in which $\appa$ is Abelian and show that the classical  case of parametric models of probability distributions endowed with the Fisher-Rao metric tensor perfectly falls in the setting we  introduced.
Finally, we discuss the case of rank-one, strongly-continuous unitary models on a complex, separable Hilbert space $\hh$.
This family of models is broad enough to encompass various models used in the quantum setting ({\itshape e.g.}, the case of normal pure states of $\bh$ endowed with the Fubini-Study metric tensor, and various notions of generalized coherent states).
Section \ref{sec: conclusions} closes the work with some thoughts and comment on possible future investigations stemming from the formalism introduced.

\section{\walgs\ and normal positive linear functionals}\label{sec: walgs and nplfs}

Before  giving the precise definition of  {\itshape smooth parametric model of normal positive linear functionals}  on a $W^{\star}$-algebra $\appa$, we have to recall some notions and some well-known results about $C^\star$-algebras and $W^\star$-algebras. 
We assume basic knowledge of functional analysis (e.g., Banach spaces, Hilbert spaces, $\mathcal{L}^{p}$-spaces), and we refer to standard introductory texts such as \cite{Blackadar-2006,B-R-1987-1,Sakai-1997,Takesaki-2002} for additional information on the subject.

\begin{definition}
A complex Banach space $(A,+,||\cdot||)$ on which there is an associative product $(\mathbf{a},\mathbf{b})\mapsto \mathbf{a}\cdot\mathbf{b}\equiv\mathbf{a}\mathbf{b}$ such that $||\mathbf{a}\mathbf{b}||\leq||\mathbf{a}||\,||\mathbf{b}||$ is called a complex {\bfseries Banach algebra}.
A complex Banach algebra $(A,+,||\cdot||,\cdot)$ on which there is a bounded, anti-linear map $\mathbf{a}\mapsto\mathbf{a}^{\dagger}\in A$ such that $(\mathbf{a}^{\dagger})^{\dagger}=\mathbf{a}$ and $(\mathbf{a}\mathbf{b})^{\dagger}=\mathbf{b}^{\dagger}\,\mathbf{a}^{\dagger}$ is called a complex {\bfseries involutive Banach algebra}.
A complex involutive Banach algebra $(A,+,||\cdot||,\cdot,\dagger)$ such that $||a||^2 = ||a^*a||$ is called a {\bfseries \calg}.
For the sake of notational simplicity, we simply write $\appa$ to denote a \calg\ $(A,+,||\cdot||,\cdot,\dagger)$. 
A {\bfseries $W^{\star}$-algebra} is a \calg\ $\appa$ which is isomorphic to the Banach dual space of a Banach space $\mathcal{B}(\appa)$ called the predual of $\appa$ (this predual space is essentially unique \cite[cor. III-3.9]{Takesaki-2002}).
Note that $\mathcal{B}(\appa)$ is uniquely determined by $\appa$.
\end{definition}

If $\appa$ admits an identity $\mathbb{I}$, it is called a unital $C^{\star}$-algebra.
Note that every \walg\ has an identity \cite[Par. 1.7]{Sakai-1997}.
An element $\mathbf{a}$ in the \calg\ $\appa$  such that $\mathbf{a}^{\dagger}=\mathbf{a}$ is called {\bfseries self-adjoint}, while an element $\mathbf{a}$   such that $\mathbf{a}^{\dagger}=-\mathbf{a}$ is called {\bfseries skew-adjoint}.
The set of self-adjoint elements in $\appa$ is denoted by $\appa_{sa}$ and it is a \grit{real} Banach subspace of the Banach space underlying $\appa$ when the latter is considered as a \grit{real} Banach space.
More importantly, every skew adjoint element $\mathbf{a} \in  \appa$ can be uniquely written as $\mathbf{a}=i\mathbf{b}$ for some $\mathbf{b}\in\appas$, and every element $\mathbf{a} \in  \appa$ can be uniquely written as $\mathbf{a}=\mathbf{b} + i\mathbf{c}$ for some $\mathbf{b},\mathbf{c}\in\appas$.
This fact implies that, as a real Banach space, $\appa$ is the direct sum of $\appas$ with $i\appas$ (the space of skew-adjoint elements), while, as a complex Banach space, $\appa$ is the complexification of $\appas$.

The set $\appas$ endowed with the anti-commutator product
\be\label{eqn: Jordan product}
\{ \mathbf{a} , \mathbf{b} \} := \frac{1}{2} ( \mathbf{a} \mathbf{b} + \mathbf{b} \mathbf{a} )
\ee
is also an important example of Banach-Jordan algebra \cite{A-S-2001}, that is, a Banach space which is also a Jordan algebra whose Jordan product is continuous in the norm topology.
Quite interestingly, $\appas$ can be turned into a Banach-Lie algebra  (a Banach space which is also a Lie algebra whose Lie product is continuous in the norm topology) with the Lie product
\be\label{eqn: Lie product}
[\mathbf{a} , \mathbf{b} ] := \frac{1}{2i} ( \mathbf{a} \mathbf{b} - \mathbf{b} \mathbf{a} ).
\ee
The Jordan product and Lie product on $\appas$ speak to each other according to the identity
\be\label{eqn: Jordan-Lie compatibility} 
\{\{\mathbf{a},\mathbf{b}\},\mathbf{c}\} - \{\mathbf{a},\{\mathbf{b},\mathbf{c}\}\}=[[\mathbf{a},\mathbf{c}],\mathbf{b}]
\ee
for all $\mathbf{a},\mathbf{b},\mathbf{c}\in\appas$.
Equation \eqref{eqn: Jordan-Lie compatibility} can be verified by a direct computation for $\appas$, and it's at the earth of the definition of  abstract  Lie-Jordan algebras \cite{F-F-M-P-2014,F-F-I-M-2013}.


What is perhaps the prototypical example of a unital \calg\ is the space $\bh$ of bounded linear operators on the complex Hilbert space $\hh$ in which the addition and multiplication are those between linear operators, the norm is the operator norm, and the involution is the map sending a linear operator in its adjoint in the Hilbert-space sense.
Accordingly, $\bhsa$ is the space of bounde self-adjoint linear operators on $\hh$, while $\bhpos$ is the space of bounded positive (non-negative) linear operators on $\hh$.
The \calg\ $\bh$ is non-commutative whenever $\mathrm{dim}(\hh)\geq 2$.
From standard results in the theory of bounded operators on Hilbert spaces it follows that $\bh$ is the dual Banach space of the space $\mathcal{T}(\hh)$ of trace-class linear operators on $\hh$.
Consequently, $\bh$ is a $W^{\star}$-algebra.
The \walg\ $\bh$ is the one most often used to describe quantum systems especially in the context of the so-called standard quantum mechanics \cite{C-I-M-M-2015,Landsman-1998,Landsman-2017}, in quantum information theory and quantum computing \cite{N-C-2011,Petz-2007}, and in quantum information geometry  \cite{A-N-2000,C-J-S-2020-02,C-J-S-2020}.

On the other hand, there are also  examples of Abelian (commutative) \calgs\ which are of paramount importance to describe classical systems, especially in the context of statistical mechanics \cite{Landsman-1998,Landsman-2017} and classical information geometry \cite{G-N-2020,G-R-2006,G-R-2006-02}.
For instance, take a  topological  space $\mathcal{X}$ and consider the space $C(\mathcal{X})$ of complex-valued continuous functions on $\mathcal{X}$.
This space is clearly an algebra with respect to the usual addition and multiplication between functions, and it admits and algebraic involution given by complex conjugation, i.e., $f^{\dagger}(x)=\overline{f(x)}$.
Without appealing to additional structures on $\mathcal{X}$, we can define a norm-like function on $C(\mathcal{X})$ given by the supremum, i.e., $||f||_{\infty}:=\sup\{|f(x)|\colon\,x\in\mathcal{X}\}$.
Clearly, unless  $\mathcal{X}$ is compact, there are functions   $f\in C(\mathcal{X})$ such that $||f||_{\infty}=\infty$.
However, both
\be
\begin{split}
C_{b}(\mathcal{X})&=\left\{f\in C(\mathcal{X})\,\colon\; ||f||_{\infty}<\infty \right\}  \\
C_{0}(\mathcal{X})=\left\{f\in C(\mathcal{X})\,\colon;\right. &\left. \forall \epsilon>0 \;\exists \mbox{ a  compact } K_{\epsilon}\subseteq\mathcal{X}\,s.t.\;|f(x)|<\epsilon\;\forall x\in K_{\epsilon}\right\}
\end{split}
\ee
are easily seen to be Banach spaces with respect to $||\cdot||_{\infty}$, and, more importantly, both $\left(C_{b}(\mathcal{X}),+,||\cdot||_{\infty},\cdot,\dagger\right)$  and $\left(C_{0}(\mathcal{X}),+,||\cdot||_{\infty},\cdot,\dagger\right)$ are  easily seen to be \calgs\ for every  topological Hausdorff space $\mathcal{X}$, denoted $C_{b}(\mathcal{X})$ and $C_{0}(\mathcal{X})$, respectively, with an evident abuse of notation.  
Quite often, the topological space $\mathcal{X}$ is taken to be locally-compact and Hausdorff.
We can motivate this choice noting that, when $\mathcal{X}$ is not Hausdorff we can always consider its Stone-Cech compactification $\beta\mathcal{X}$ and $C_{b}(\mathcal{X})\cong C_{b}(\beta\mathcal{X})\equiv C(\beta\mathcal{X})$, and when $\mathcal{X}$ is not locally-compact  then $C_{0}(\mathcal{X})$ may be “very little” if not empty (e.g., when $\mathcal{X}$ is an infinite dimensional Banach space because by Riesz lemma every $K_{\epsilon}$ has empty interior).

A relevant example of Abelian \walg\ is borrowed from measure theory.
Specifically, let $\mathcal{X}=(X,\Sigma)$ be a measurable space and let $\mu$ be a measure on it.
Let $\mathcal{F}(\mathcal{X})$ denote the space of complex-valued measurable functions on $\mathcal{X}$, define
\be
L^{\infty}(\mathcal{X},\mu)=\left\{f\in\mathcal{F}(\mathcal{X})\,\colon\; \exists M \geq 0\;s.t.\;|f(x)| < M \,\mu-a.e. \right\}.
\ee
Clearly, we can endow $L^{\infty}(\mathcal{X},\mu)$ with the usual pointwise product among functions, and with the involution $\dagger$ associated with complex conjugation as it is done for continuous functions.
Concerning the norm, if we set $||f||_{\infty}:=\inf \{C\geq 0\,\colon |f(x)|\leq C\,\mu-a.e.\}$ for every $f\in L^{\infty}(\mathcal{X},\mu)$, we have that $N=ker(||\cdot||_{\infty})\neq\emptyset$.
However, defining all algebraic operations “by projection”, it is an instructive exercise to prove that $L^{\infty}(\mathcal{X},\mu)=\left(L^{\infty}(\mathcal{X},\mu)/N,+,||\cdot||_{\infty}\right)$ is a complex Banach space and that $\left(L^{\infty}(\mathcal{X},\mu),\cdot, \dagger\right)$ is a  is a $C^{\star}$-algebra which we is denoted by $L^{\infty}(\mathcal{X},\mu)$ with an evident abuse of notation.
Then, the standard theory of $\mathcal{L}^{p}$-spaces guarantees that $\mathcal{L}^{\infty}(\mathcal{X},\mu)$ is the Banach space dual of $\mathcal{L}^{1}(\mathcal{X},\mu)$ so that $\mathcal{L}^{\infty}(\mathcal{X},\mu)$ is indeed an Abelian $W^{\star}$-algebra.

\vsp

Let us now turn our attention from \calgs\ and \walgs\ to positive linear functionals on them.
As the name suggest, they are elements of the Banach dual space $\appa^{\ast}$ of the \calg\ (or \walg) $\appa$, but of a very special kind.
Indeed, they basically formalize and generalize the notion of positive measure from measure theory to the framework of \calgs\ and $W^{\star}$-algebras.
From a more practical point of view, a positive linear functional on $\appa$ is an element of $\appas^{\ast}\subset\appa^{\ast}$, that is, a bounded linear functional which takes real values on self-adjoint elements (these functionals are also called \grit{self-adjoint}).
Moreover, as the name already suggests, they satisfy a particular positivity property.
In order to state this property, we first introduce the notion of {\bfseries positive} element in $\appa$.
An element  $\mathbf{a}\in\appa$ such that $\mathbf{a}=\mathbf{b}^{\dagger}\mathbf{b}$ for some $\mathbf{b}\in\appa$ called {\bfseries positive}, and it is clear that every positive element is also self-adjoint.
The space of positive elements in $\appa$ is denoted as $\appa_{+}$ and is a norm-closed convex cone inside $\appa_{sa}$.

In the classical cases discussed above, it is immediate to check that positive elements correspond to positive functions (or equivalence classes of positive functions), while, in the quantum case, they correspond to bounded non-negative linear operators, that is, all those bounded linear operators whose spectrum lies on the positive semiline (actually, the notion of spectrum also makes sense for elements in a \calg\ or \walg\ in such a way that positive elements are precisely those elements whose specturm lies on the positive semiline as it happens for bounded linear operators).

Then the set $\mathcal{P}$ of positive linear functionals on $\appa$ is defined as
\begin{equation}
\mathcal{P} := \{ \omega \in \appas^{\ast} \, | \, \omega(\mathbf{p}) \geq 0 \quad \forall \, \mathbf{p} \in \mathscr{A}_+ \}.
\end{equation}
It is almost trivial to check that $\mathcal{P}$ is a convex cone in $\appas^{\ast}\subset\appa^{\ast}$.
A positive linear functional $\omega$ is called \grit{faithful} if $\omega(\mathbf{a}^{\dagger}\mathbf{a})=0$ is equivalent to $\mathbf{a}=\mathbf{0}$.
The space of faithful positive linear functionals is often denoted by $\mathcal{P}_{+}$.
A positive linear functional with unit norm is called a \grit{state}, and the space of (faithful) states on $\appa$ is denoted by $\mathcal{S}$ ($\mathcal{S}_{+}$).
Clearly, $\mathcal{S}$ is a convex set, and it is weakly*-compact if and only if $\appa$ has an identity element \cite[Thm. 2.3.15]{B-R-1987-1}.
In this case, $\mathcal{S}$ is the weak*-closure of the convex envelope of the space of pure states on $\appa$.
A state $\rho$ is defined to be \grit{pure} if, given $\omega\in\mathcal{P}$, it holds that $\rho -\omega$ is a positive linear functional if and only if $\omega=\lambda\rho$ with $0\leq \lambda\leq 1$.

Positive linear functionals on $\appa$ posses particularly nice properties: 
\begin{itemize}
\item\label{plfs 1} $\omega(\mathbf{a}^{\dagger} )=\overline{\omega( \mathbf{a})}$ for all $\mathbf{a} \in\appa$;
\item\label{plfs 2} $|\omega(\mathbf{a}^{\dagger}\mathbf{b})|^{2}\leq \omega(\mathbf{a}^{\dagger}\mathbf{a})\omega(\mathbf{b}^{\dagger}\mathbf{b})$ for all $\mathbf{a},\mathbf{b}\in\appa$ \cite[lem. 2.3.10]{B-R-1987-1};
\item\label{plfs 3} if $\appa$ has an identity $\mathbb{I}$, then $||\omega||=\omega(\mathbb{I})$ for all $\omega\in\mathcal{P}$ \cite[prop. 2.3.11]{B-R-1987-1};
\item\label{plfs 4}  $||\omega + \sigma||=||\omega|| + ||\sigma||$ for all $\omega,\sigma\in\mathcal{P}$ \cite[cor. 2.3.12]{B-R-1987-1}.
\end{itemize}

When $\appa$ is a \walg\ we may single out an interesting family of elements in $\pos$ called \grit{normal}.
Specifically, we recall that $\appa$ is isomorphic to the Banach dual space of its predual space $\mathcal{B}$, and thus there is the canonical immersion of $\mathcal{B}$ into its double dual $\mathcal{B}^{\ast\ast}\cong \appa^{\ast}$ given by $\mathbf{x}\mapsto \xi_{\mathbf{x}}$ with $\xi_{\mathbf{x}}(\mathbf{a})=\mathbf{a}(\mathbf{x})$ where $\mathbf{a}$ is identified with an element of the dual of $\mathcal{B}$.
We denote by $\appa_{\ast}$ the norm-closed image of the predual space $\mathcal{B}$ through the canonical immersion, and all those linear functionals in $\appa_{\ast}$ are called \grit{normal}.
From a topological point of view, the set of normal linear functionals $\appa_{\ast}$ on $\appa$ is the set of all norm-continuous linear functionals on $\appa$ which are continuous also with respect to the weak topology on $\appa$ determined by the predual space $\mathcal{B}$ \cite[thm. 1.13.2, p. 28]{Sakai-1997}.
This weak topology on $\appa$ is referred to as the $\sigma$-weak (or ultraweak) topology on $\appa$.
Quite obviously, the normal positive linear functionals (\nplfs)  are just elements of the set $\pos=\mathcal{P}\cap\mathscr{A}_{\ast}$, while normal states are elements of $\stsp=\mathcal{S}\cap\appa_{\ast}$.
It turns out that $\omega\in\mathcal{P}$ being normal is equivalent to the validity of the equality
\be
\omega\left(\sum_{j\in\mathrm{J}}\mathbf{e}_{j}\right)=\sum_{j\in\mathrm{J}}\omega\left(\mathbf{e}_{j}\right)
\ee
for every orthogonal family $\{\mathbf{e}\}_{j\in\mathrm{J}}$ of projections in $\appa$ \cite[cor. III.3.11, p. 136]{Takesaki-2002}.



\vsp

Going back to the quantum case $\appa=\bh$, we recall that $\appa$ is the dual space of the space $\mathcal{T}(\hh)$ of trace-class linear operators on $\hh$.
The canonical immersion of $\mathcal{T}(\hh)$ into $\appa^{\ast}=(\bh)^{\ast}$ is given by $\tilde{\xi}\mapsto\xi $ with
\be
\xi(\mathbf{a})=\mathrm{Tr}_{\hh}(\tilde{\xi}\mathbf{a})
\ee
for all $\mathbf{a}\in\appa=\bh$.
Therefore, the space $\pos$ of \nplfs\ on $\appa$ is identified with the space of positive trace-class linear operators $\tilde{\rho}$ on $\hh$, and the space $\stsp$ of normal states on $\appa$ is identified with the convex subset of $\pos$ singled out by the constraint $\mathrm{Tr}_{\hh}(\tilde{\rho})=1$.

On the other hand, in the classical case $\appa=L^{\infty}(\mathcal{X},\mu)$, the \walg\ $\appa$ is the dual space of $L^{1}(\mathcal{X},\mu)$ so that we may identify a normal linear functional $\xi\in\appa_{\ast}$ with the complex measure $\mu_{\xi}$ given by
\be\label{eqn: normal linear functionals as complex measures in the Abelian case}
\mu_{\xi}(A)=\int_{A}\,f_{\xi}\,\mathrm{d}\mu
\ee
for every measurable subset $A\subseteq \mathcal{A}$, with $f_{\xi}\in L^{1}(\mathcal{X},\mu)$.
Accordingly, the space $\pos$ of \nplfs\ on $\appa=L^{\infty}(\mathcal{X},\mu)$ is identified with the space of positive $f\in L^{1}(\mathcal{X},\mu)$, and the space $\stsp$ of normal states with the space of probability density functions in $L^{1}(\mathcal{X},\mu)$.

\vsp

From the practical point of view, it turns out that most of the parametric models used in classical and quantum information geometry  consist of \nplfs\  when framed in the context of \walgs\ as we do in this work.
This instance may be partly attributed to the nice mathematical properties of \nplfs.
Be that as it may, motivated by the above-mentioned empirical usefulness of \nplfs\ with respect to classical and quantum information geometry,  we decided to limit our analysis to this kind of positive linear functionals and leave a more general analysis for future works.

Given an arbitrary \nplf\ $\omega\in\pos$, there exists a unique non-zero projection $\mathbf{p}\in\appa$ (called the \grit{support projection} of $\omega$) such that 
\be\label{eqn: nplf and its support projection}
\omega(\mathbf{a})=\omega(\mathbf{ap})=\omega(\mathbf{pa})= \omega(\mathbf{pap})
\ee
for all $\mathbf{a}\in\appa$, and such that $\omega$ is faithful when restricted to $\appa_{pp}=\mathbf{p}\appa\mathbf{p}$ \cite[lem. III.3.6, p. 134]{Takesaki-2002}.
Given the support projection $\mathbf{p}$ of $\omega$, we have the topological direct sum decomposition
\be\label{eqn: decomposition of A wrt to the support proj of a nplf}
\appa=\appa_{pp}\oplus\appa_{pq}\oplus\appa_{qp}\oplus\appa_{qq},
\ee
where the topology we are referring to is  the one induced by the $C^{\star}$-norm, where $\mathbf{q}=\mathbb{I}-\mathbf{p}$, and where we have set $\appa_{pp}=\mathbf{p}\appa\mathbf{p},$ $\appa_{pq}=\mathbf{p}\appa\mathbf{q},$ $\appa_{qp}=\mathbf{q}\appa\mathbf{p},$ and $\appa_{qq}=\mathbf{q}\appa\mathbf{q}$.
We often refer to $\appa_{pp}$ as the \grit{support algebra} of $\omega$.
Note that $\appa_{pp}$ and $\appa_{qq}$ are \calgs\ and that the involution $\dagger$ gives a Banach space isomorphism between $\appa_{qp}$ and $\appa_{pq}$.
Obviously, if $\omega$ is faithful, then $\mathbf{p}=\mathbb{I}$ so that $\appa_{pp}=\appa$ and all other summands in the right hand side of equation \eqref{eqn: decomposition of A wrt to the support proj of a nplf} vanish.

\begin{remark}\label{rem: matrix-like calculus}
The decomposition in equation   \eqref{eqn: decomposition of A wrt to the support proj of a nplf} allows us to visualize the algebraic operations in $\appa$ in terms of matrix operation.
Specifically, given an arbitrary $\mathbf{a}\in\appa$, if we write
\be
\mathbf{a}=\begin{pmatrix}  \mathbf{a}_{pp}= \mathbf{p} \mathbf{a} \mathbf{p} & \mathbf{a}_{pq} =\mathbf{p} \mathbf{a} \mathbf{q}\\ \mathbf{a}_{qp} = \mathbf{q} \mathbf{a} \mathbf{p}& \mathbf{a}_{qq} = \mathbf{q} \mathbf{a} \mathbf{q}\end{pmatrix},
\ee
it is a matter of direct inspection to check that $\mathbf{a} + \mathbf{b}$ and $\mathbf{ab}$ can be computed using matrix-like operations.
It then follows that the so-called Gel'fand ideal  $N_{\omega}$ of $\omega$ defined by
\be\label{eqn: Gelfand ideal}
N_{\omega}:=\{\mathbf{a}\in\appa\colon\;\;  \omega(\mathbf{a}^{\dagger}\mathbf{a})=0\}
\ee
can be written as
\be\label{eqn: Gelfand ideal nplf}
N_{\omega} = \appa_{pq}\oplus\appa_{qq}.
\ee
The Gel'fand ideal is instrumental to the construction of the so-called Gel'fand-Naimark-Segal (GNS) representation of $\appa$ associated with $\omega$ \cite[Sec. 2.3]{B-R-1987-1}.
\end{remark}

\section{Parametric models of normal positive linear functionals}\label{sec: parametric models}

Once all the technical tools discussed in section \ref{sec: walgs and nplfs} are at our disposal, we are ready to define  what a {\itshape  smooth  parametric model of} \nplfs\ actually is.

\begin{definition}\label{defn: parametric models}
Given a $W^{\star}$-algebra $\appa$, a smooth parametric model  of \nplfs\ is a triple $(\mathcal{M}, j, \mathscr{A})$ where $\mathcal{M}$ is a Banach manifold  and $j: \mathcal{M} \longrightarrow (\appa_{sa})^{\ast}$ is a smooth map such that $j(\mathcal{M}) \subset \mathscr{P}$. 
If $j(\mathcal{M})\subset \stsp\subset \mathscr{P}$, we refer to $(\mathcal{M}, j, \mathscr{A})$ as a smooth parametric model  of normal states.
A  smooth parametric model  of \nplfs\  (normal states) $(\mathcal{M}, j, \mathscr{A})$ is said to be \textbf{identifiable} if $j$ is injective, and it is said to be \textbf{locally identifiable} if $T_m j$ is injective for all $m \in \mathcal{M}$.
\end{definition}

\begin{remark}
In the following in order to avoid burdensome repetitions of words, we  often write “parametric model” when we actually means “smooth parametric model  of \nplfs\ (normal states)”.
\end{remark}

Clearly, every identifiable model is locally identifiable while the converse is not true.
Indeed, if  $(\mathcal{M}, j, \mathscr{A})$ is locally identifiable, for every point $m \in \mathcal{M}$, the best we can do is find  an open neighbourhood $U$ of $m$ such that the model is identifiable when restricted to $U$. 

If $\pos$ were a  Banach manifold it would clearly admit a tangent space at each point and the tangent map  of the immersion map $\jj$ associated with a parametric model $(\mathcal{M}, j, \mathscr{A})$ would send a tangent vector at $m\in \mathcal{M}$ into a tangent vector lying in the tangent space of $\pos$ at $\jj(m)\in\pos$.
Since $\pos$ is not a Banach manifold (not even in the finite-dimensional case), we can no longer use the tangent space of $\pos$.
However, being $\pos$ a subset of the Banach space $\appa^{\ast}_{sa}$ there is a kind of “close relative” of the tangent space we can exploit:  the tangent double tangent at $\pos$.

\begin{definition}[Tangent Double Cone]\label{def:tangent-double-cone}
Let $\mathcal{B}$ be a real Banach space, $X$ a subset of $\mathcal{B}$ and let $\xi \in X$. 
Then, $\eta \in \mathcal{B}$ is said to be in the tangent double cone $T_{\xi} X$ of $X$ at $\xi$ if there exist $\epsilon > 0$ and a $C^1$-curve $\gamma:(-\epsilon,\epsilon) \longrightarrow \mathcal{B}$ s.t.
\begin{equation}
\gamma(t) \in X; \quad \forall t \in (-\epsilon,\epsilon); \quad \gamma(0) = \xi; \quad \dot{\gamma}(0)=\eta.
\end{equation}
That is, if there exist a curve that has $\eta$ as derivative at the point $\xi$.
\end{definition}

\begin{remark}
The tangent double cone reduces to the ordinary tangent space whenever the subset $X$ considered in Definition \ref{def:tangent-double-cone} is actually a submanifold of $\mathcal{B}$.
 
\end{remark}

With respect to the notion of parametric model introduced in definition \ref{defn: parametric models}, the relevance of the tangent double cone $T_{\omega}\pos$ at every \nplf\ $\omega$ stems from the following proposition whose proof follows directly from the definition of tangent vector and tangent map at a point.

\begin{proposition}\label{prop: tangent space of parametric model immersed in tangent double cone}
Let $(\mathcal{M},\jj,\appa)$ be a smooth parametric model of \nplfs\ according to definition \ref{defn: parametric models}.
Then, $T_{m}\jj(T_{m}M)$ lies in the tangent double cone $T_{\jj(m)}\pos$ for all $m\in\mathcal{M}$.
\end{proposition}

 
Motivated by proposition \ref{prop: tangent space of parametric model immersed in tangent double cone}, we investigate  the structure of $T_{\omega}\pos$.
A first interesting fact about $T_{\omega}\pos$ is that it is actually a vector space containing only normal linear functionals.

\begin{proposition}  \label{prop:double_tangent_cone_normal}
Let $\mathcal{B}$ be a real Banach space, let $X\subseteq \mathcal{B}$ let  $\xi \in X$, and  let $T_{\xi}X$ be the tangent double cone of $X$ at $\xi$.
Then, if $X$ is a convex cone,  $T_{\xi}X$ is a vector space.
On the other hand, if there is a Banach subspace $\mathcal{V}$ of $\mathcal{B}$ such that $X\subseteq\mathcal{V}$, then $T_{\xi}X\subseteq \mathcal{V}$.
In particular, the tangent double cone $T_{\omega}\pos$ is a vector space for every $\omega\in\pos$ sitting inside $(\appa_{sa})_{\ast}$ for every $\omega\in\pos$.
\end{proposition}

\proof

Suppose $X$ is a convex cone.
Let $\eta,\zeta$ be in $T_\xi X$, and let $\gamma$ and $\sigma$ be the smooth curves in $X$ starting at $\xi$ and having $\eta$ and $\zeta$ as tangent vectors at $t=0$, respectively. 
Let $I_{\gamma}=(-\epsilon_{\gamma},\epsilon_{\gamma})$ be the domain of $\gamma$.
Let $a\in\mathbb{R}$.
If $a=0$, then we define $I_{a\gamma}=(-\epsilon_{\gamma},\epsilon_{\gamma})$ as the domain of the curve $\gamma_{a}(t)=\gamma(0)$ starting at $\xi$ and having $a\eta=0$ as the tangent vector  at $t=0$.
If $a>0$, then we define $I_{a\gamma}=(-\epsilon_{\gamma}/a,\epsilon_{\gamma}/a)$ as the domain of the curve $\gamma_{a}(t)=\gamma(at)$ starting at $\xi$ and having $a\eta $ as the tangent vector  at $t=0$.
If $a<0$, then we define $I_{a\gamma}=( \epsilon_{\gamma}/a,-\epsilon_{\gamma}/a)$ as the domain of the curve $\gamma_{a}(t)=\gamma(at)$ starting at $\xi$ and having $a\eta $ as the tangent vector at $t=0$.
Similar considerations apply for $I_{\sigma}$.
Then, take $a,b\in\mathbb{R}$ with $a,b>0$, $0<\alpha<1$,   $\beta=(1-\alpha)$, and $I_{\mu}=(-\epsilon_{\mu},\epsilon_{\mu})$ with $\epsilon_{\mu}=\mathrm{min}(\alpha/a, \beta/b)$,  and define the curve $\mu\colon I_{\mu}\ra  \mathcal{B}$ setting $\mu(t):= \alpha\gamma(at/\alpha) + \beta\sigma(bt/\beta)$.
Being $X$ a convex cone, $\mu(t)$ lies in $X$ for every $t\in I_{\mu}$, and it is a matter of direct inspection to check that $\mu$ is a smooth curve starting at $\xi$ having $a\eta + b\zeta$ as tangent vector at $t=0$.
All other cases in which $a$ and $b$ are either positive, negative or $0$ are handled similarly.
It then  follows that  $T_{\xi}X$ is a vector space as claimed.
Now,  assume there is a Banach subspace $\mathcal{V}$ of $\mathcal{B}$ such that $X\subseteq\mathcal{V}$.
Let $\eta$ be in $T_\xi X$, this means that there exists a curve $\gamma(t)$ such that $\gamma(0) = \xi$ and $\dot{\gamma}(0) = \eta$, meaning that
\begin{equation} \label{eqn:derivative_proof_prop_1}
\eta = \lim_{\epsilon \to 0} \frac{\gamma(\epsilon) - \gamma(0)}{\epsilon}.
\end{equation}
Since $X\subseteq\mathcal{V}$ and $\mathcal{V}$ is a vector space, it is clear that $\gamma(\epsilon)$, $\gamma(0)$, $\gamma(\epsilon)-\gamma(0)$ are in $\mathcal{V}$
Therefore, since the limit in equation \eqref{eqn:derivative_proof_prop_1} is taken in the norm topology of $\mathcal{B}$ and this topology  coincides with the norm topology on $\mathcal{V}$ because the latter is a Banach subspace of the former, we conclude that $\eta\in\mathcal{V}$ as claimed.
Eventually, since $\pos$ is a convex cone inside $(\appa_{sa})_{\ast}\subseteq\appa_{sa}^{\ast}$  and $(\appa_{sa})_{\ast}$ is a Banach subspace of $\appa_{sa}^{\ast}$ \cite[p. 29]{Sakai-1997}, we conclude that $T_{\omega}\pos$ sitting inside $(\appa_{sa})_{\ast}$ for every $\omega\in\pos$.
\qed


\begin{remark}\label{rem: classical tangent double cone is closed}
In the Classical case $\appa=L^{\infty}(\mathcal{X},\mu)$, the tangent double cone at each \nplf\ $\omega$ is closed and coincides with the space of signed measures absolutely continuous with respect to $\mu_{\omega}$ \cite[thm. 3.1, p. 42]{A-J-L-S-2017}.
Therefore, the tangent double cone is isomorphic with the Banach space $ L^{1}(\mathcal{X},\mu_{\omega})$.

\end{remark}

Quite interestingly, it turns out there is a suitable notion of absolute continuity in the context of bounded linear functionals on a \calg\ $\appa$ \cite{Niestegge-1983}  generalizing the measure-theoretic notion of absolute continuity to the $C^{\ast}$-algebraic case, and such that, when $\appa$ is a \walg, the norm-closure of the tangent double cone $T_{\omega}\pos$ coincides precisely with the space of bounded linear functionals which are absolutely continuous with respect to the \nplf\ $\omega$ in perfect analogy with what happens in the classical case recalled in remark \ref{rem: classical tangent double cone is closed}.

\begin{definition}[Absolute continuity]\label{def:absolute-continuity}
Let $\omega$ be a positive functional on the \calg\ $\appa$, $\eta \in (\mathscr{A}_{sa})^*$ and let also $\mathbf{B}_1$ be the unit ball in $\mathscr{A}$. Then $\eta$ is said to be absolutely continuous with respect to $\omega$ if one of the following equivalent statements is true:
\begin{itemize}
\item $\forall \epsilon > 0, \exists \, \delta > 0 \quad s.t. \quad \omega(\mathbf{a}) < \delta \implies \left| \eta(\mathbf{a}) \right| < \epsilon  \quad \forall \, \mathbf{a} \in \mathscr{A}_+\cap \mathbf{B}_1$.
\item For every sequence $\{\mathbf{a}_n\}_{n\in\mathbb{N}}$ with $\mathbf{a}_n \in \mathscr{A}_+\cap \mathbf{B}_1$, then $\lim \omega(\mathbf{a}_n) = 0 \implies \lim \eta(\mathbf{a}_n) = 0$ .
\item For every net $\{\mathbf{a}_\lambda\}_{\lambda\in\Lambda}$ with $\mathbf{a}_\lambda \in \mathscr{A}_+\cap \mathbf{B}_1$, then $\lim \omega(\mathbf{a}_\lambda) = 0 \implies \lim \eta(\mathbf{a}_\lambda) = 0$.
\end{itemize}
\end{definition}

We will denote the set of all self-adjoint linear functionals that are absolutely continuous with respect to a positive linear functional $\omega$ by $AC_{\omega}$.
We now list some useful properties of $AC_{\omega}$:
\begin{enumerate}
\item\label{point: AComega and Jomega} for every \calg\ $\appa$ and every positive linear functional $\omega$ on $\appa$, the set $AC_{\omega}$ is a Banach subspace of $(\appas)^{\ast}$ which is the  closure of 
\be\label{eqn: Jomega}
J_{\omega} = \bigg{\{} \eta \in (\appas)^{\ast} \, | \, \exists \, \mathbf{a} \in \appas \, : \, \eta (\mathbf{b}) =   \omega( \{ \mathbf{ a } , \mathbf{ b } \} ) \,\, \forall \mathbf{b} \in \appa \bigg{\}}
\ee
in the norm topology \cite[Thm. 1.3]{M-N-1987};
\item when $\appa$ is a \walg\ and $\omega$ is a \nplf\ on $\appa$, every $\xi\in J_{\omega}$ is normal \cite[Sec. 1.8]{Sakai-1997}, and thus, since $\appa_{\ast}$ is norm-closed in $\appa^{\ast}$, point \ref{point: AComega and Jomega} implies that $AC_{\omega}\subseteq (\appas)_{\ast}$;
\item\label{point: AComega for W*} when $\appa$ is a \walg\ and $\omega$ is a \nplf, then $\xi\in(\appas)_{\ast}$ is in $AC_{\omega}$ iff $\omega(\mathbf{p})=0$ implies $\xi(\mathbf{p})=0$ for every projection $\mathbf{p}\in\appa$ \cite[Prop. 1.5]{Niestegge-1983};
\item\label{point: AComega for C0} in the Classical case $\appa=C_{0}(\mathcal{X})$, $\xi\in(\appas)^{\ast}$ is in $AC_{\omega}$ iff $\mu_{\xi}$ is absolutely continuous with respect to $\mu_{\omega}$ in the measure-theoretic sense \cite[Prop. 1.3]{Niestegge-1983};
\item\label{point: AComega for Linfinity} in the Classical case $\appa=L^{\infty}(\mathcal{X},\mu)$, projections in $\appa$ are in one-to-one correspondence with characteristic functions on measurable subsets of $\mathcal{X}$ with finite $\mu$-measure, so that the result recalled in point \ref{point: AComega for W*} above is equivalent to $\mu_{\xi}$ being absolutely continuous with respect to $\mu_{\omega}$ in the measure-theoretic sense.
\end{enumerate}

The set $J_{\omega}$ in equation \eqref{eqn: Jomega} turns out to be particularly useful for our investigation, and we  write a generic element in $J_{\omega}$ as $\eta_{\mathbf{a}}$ with $\mathbf{a}\in\appas$ to emphasize the intimate connection between elements in $J_{\omega}$ and elements in $\appas$ given by
\be\label{eqn: elements in Jomega}
\eta_{\mathbf{a}}(\mathbf{b})=\omega(\{\mathbf{a},\mathbf{b}\})
\ee
for all $\mathbf{b}\in\appas$.

\begin{proposition}\label{prop: norm-closure of tangent double cone}
The norm-closure $\overline{T_{\omega}\pos}$ of the tangent double con $T_{\omega}\pos$ at $\omega\in\pos$ coincides with $AC_{\omega}$.
\end{proposition}
\proof
First, we prove that $AC_{\omega}\subseteq \overline{T_{\omega}\pos}$.
Consider then the smooth curve $\omega_t$ in $(\appas)_{\ast}$ defined by
\begin{equation}
\omega_t(\mathbf{b})= \omega( e^{\frac{t \mathbf{a}}{2}} \mathbf{b} \, e^{\frac{t \mathbf{a}}{2}} ).
\end{equation}
Clearly,   $\omega_0 = \omega$ and $\omega_{t}\in\pos$ for all $t\in\mathbb{R}$ so that the tangent vector at $t=0$ is in $T_\omega\mathscr{P}$ by definition of the tangent double cone. 
A quick computation shows that the tanget vector at $t=0$ reads
\begin{equation}
\dot{\omega_t}( \mathbf{b}) \big|_{t=0} = \frac{1}{2} \, \omega(\mathbf{a} \mathbf{b} + \mathbf{b} \mathbf{a}) = \eta_{\mathbf{a}}(\mathbf{b}),
\end{equation}
with $\eta_{\mathbf{a}}\in J_{\omega}$ because of equation \eqref{eqn: elements in Jomega}, and it follows that   $J_\omega \subseteq T_\omega \mathscr{P}$ so that $AC_\omega = \overline{J_{\omega}} \subseteq \overline{T_\omega \mathscr{P}}$ (recall point \ref{point: AComega and Jomega}).

We now prove that $ \overline{T_{\omega}\pos} \subseteq AC_{\omega}$.
Indeed, suppose $\xi\in T_{\omega}\pos$ and let $\gamma\colon (-\epsilon,\epsilon)\ra(\appa_{sa})_{\ast}$ be a smooth curve in $\pos$ starting at $\omega$ with $\xi$ as  tangent vector at $t=0$.
Then, for every projection $\mathbf{p}\in\appa$ such that $\omega(\mathbf{p})=0$, the function $f(t)=\omega_{t}(\mathbf{p})=(\gamma(t))(\mathbf{p})$ is a real-valued smooth function  which is non-negative and vanishes when $t=0$ so that its derivative $f'(0)$ at $t=0$ vanishes.
Direct inspection shows that $f'(0)=\xi(\mathbf{p})=$ so that $\xi\in T_{\omega}\pos$ implies $\xi\in AC_{\omega}$ because of point \ref{point: AComega for W*}, and thus $\overline{T_{\omega}\pos} \subseteq AC_{\omega}$.

\qed

\begin{proposition}\label{prop: the dual of the tangent double cone}

Given a \nplf\ $\omega$, the Banach space dual of $AC_{\omega}=\overline{J_{\omega}}$ can be identified with $\appas^{\omega}:=(\appa_{pp}\oplus\appa_{pq}\oplus\appa_{qp})_{sa}$.
\end{proposition}
\proof

Since $AC_{\omega}\subseteq (\appas)_{\ast}$ and $\appas\cong((\appas)_{\ast})^{\ast}$, it follows that  
\be
(AC_{\omega})^{\ast}\cong \appas / \mathrm{Ann}(AC_{\omega})
\ee
where 
\be
\mathrm{Ann}(AC_{\omega})=\{\mathbf{a}\in\appas \colon \eta(\mathbf{a})=0\quad\forall \eta\in AC_{\omega}\}.
\ee
If $\mathbf{a}\in \mathrm{Ann}(AC_{\omega})$ then $0=\eta_{\mathbf{a}}(\mathbf{a})=\omega(\mathbf{a}^{2})$, which means that $\mathbf{a}$ lies in the Gel'fand ideal of $\omega $ \cfr{equation \eqref{eqn: Gelfand ideal}}.
Therefore, being $\omega$ a \nplf\ and being $\mathbf{a}$ self-adjoint, equation \eqref{eqn: Gelfand ideal nplf} immediately implies that $\mathrm{Ann}(AC_{\omega})= (\appa_{qq})_{sa}$ so that $(AC_{\omega})^{\ast} \cong \appas / (\appa_{qq})_{sa} \cong (\appa_{pp}\oplus\appa_{pq}\oplus\appa_{qp})_{sa}=: \appas^{\omega} $ as claimed.

\qed

%
%

\section{J-regular parametric models}\label{sec: jordan parametric models}

Now, we want to investigate the possibility of introducing additional mathematical structures on a given {\itshape  smooth  parametric model  of} \nplfs\ (normal states).
Of course, since the parametric models  we refer to are defined in the context of $W^{\ast}$-algebras, it is only reasonable to look for the additional structures mentioned before in terms of the algebraic structures of $W^{\ast}$-algebras.
In particular, we will exploit the Jordan product among self-adjoint elements associated with the anti-commutator as in equation \eqref{eqn: Jordan product} in order to induce a (possibly degenerate) Riemannian metric tensor on suitable parametric models.

Consider a \nplf\ $\omega\in\pos $ and define the map $G_{\omega}\colon J_{\omega}\times J_{\omega}\ra\mathbb{R}$ given by
\be\label{eqn: FRBH inner product}
G_{\omega}(\eta_{\mathbf{a}},\eta_{\mathbf{b}}):= \omega(\{\mathbf{a} ,\mathbf{b}\}),
\ee
where $\eta_{\mathbf{a}},\eta_{\mathbf{b}}$ are as in equation \eqref{eqn: elements in Jomega}.
Note that $G_{\omega}$ is clearly bilinear, and it is symmetric because the Jordan product $\{,\}$ is symmetric.
Moreover, recalling equation \eqref{eqn: elements in Jomega} and equation \eqref{eqn: Gelfand ideal}, it follows that $\eta_{\mathbf{a}}$ such that
\be\label{eqn: pre-Hilbert product on Jomega}
G_{\omega}(\eta_{\mathbf{a}},\eta_{\mathbf{a}})=\omega( \mathbf{a}^{2})=0
\ee
implies $\mathbf{a}$ is in the Gel'fand ideal of $\omega$.
Therefore, being $\mathbf{a}$ self-adjoint, equation \eqref{eqn: Gelfand ideal nplf} implies that  $\mathbf{a}\in \appa_{qq}$ which means that $\eta_{\mathbf{a}}(\mathbf{b})=\omega(\{\mathbf{a},\mathbf{b}\})=0$ for all $\mathbf{b}\in\appa$  because of equation \eqref{eqn: nplf and its support projection} and equation \eqref{eqn: decomposition of A wrt to the support proj of a nplf}.
We thus conclude that $\eta_{\mathbf{a}}=\mathbf{0}$ and thus $G_{\omega}$ is positive definite on each $J_{\omega} $ for every \nplf\ $\omega$, and thus $J_{\omega}$ endowed with $G_{\omega}$ becomes a real pre-Hilbert space.

In general, $J_{\omega}$ is not closed with respect neither to the norm topology inherited from $(\appas)_{\ast}$ nor from the topology determined by the inner product given by equation \eqref{eqn: FRBH inner product}.
According to point \ref{point: AComega and Jomega}, the norm closure of $J_{\omega}$ is the space $AC_{\omega}$ of self-adjoint normal linear functionals which are absolutely continuous with respect to $\omega$.
It turns out that the closure of $J_{\omega}$ with respect to the topology determined by the inner product given by equation \eqref{eqn: FRBH inner product} (which is a real Hilbert space) can be identified with a subset of $AC_{\omega}$.

\begin{definition}\label{defn: Jordan Hilbert space}
The closure of $J_{\omega}$ with respect to the topology determined by the inner product $G_{\omega}$ \cfr{equation \eqref{eqn: FRBH inner product}} is denoted by $\mathcal{J}_{\omega}$ and it is referred to as the \grit{J-Hilbert space} of $\omega$.
\end{definition}

\begin{proposition}\label{prop: properties of Jordan Hilbert space and its dual}
Given a \nplf\ $\omega$ and the real Hilbert space $ \mathcal{J}_{\omega}$ of definition \ref{defn: Jordan Hilbert space}, it holds $\mathcal{J}_{\omega}\subseteq AC_{\omega} $ as sets.
Moreover, the dual space $(AC_{\omega})^{\ast}\cong \appas^{\omega}$ \cfr{proposition \ref{prop: the dual of the tangent double cone}} can be identified with a dense subset of the dual of $\mathcal{J}_{\omega}$.

\end{proposition}

\proof
Given $\eta\in \mathcal{J}_{\omega}$, we define a linear function $l_{\eta}\colon \appas\ra\mathbb{R}$ setting
\be
l_{\eta}(\mathbf{a}):=G_{\omega}(\eta,\eta_{\mathbf{a}})
\ee
for every $\mathbf{a}\in\appas$.
Then, exploiting the Cauchy-Schwarz inequality, it follows that
\be
|l_{\eta}(\mathbf{a})|^{2}= \left| G_{\omega}(\eta,\eta_{\mathbf{a}})\right|^{2} \leq G_{\omega}(\eta,\eta)\, G_{\omega}(\eta_{\mathbf{a}},\eta_{\mathbf{a}}) = ||\eta ||_{\omega}^{2} \,\omega(\mathbf{a}^{2})\leq  ||\eta ||_{\omega}^{2} \,||\omega || \,||\mathbf{a}||^{2} ,
\ee
which means that $l_{\eta}$ is a bounded linear map and thus defines an element in $\appas^{\ast}$.
With an evident abuse of notation, the element in $\appas^{\ast}$ determined by $l_{\eta}$ is denoted by $\eta$.
Next, it holds
\be
|\eta_{\mathbf{a}}(\mathbf{b})|^{2}\stackrel{\mbox{point \ref{plfs 2}}}{\leq} \omega(\mathbf{a}^{2})\;\omega(\mathbf{b}^{2}) \leq G_{\omega}(\eta_{\mathbf{a}},\eta_{\mathbf{a}})\;||\omega || \; ||\mathbf{b}||^{2}
\ee
which means that the norm determined by the inner product on $J_{\omega}$ is stronger than the norm on $J_{\omega}$ inherited from $(\appas)_{\ast}$.
It thus follows that $\mathcal{J}_{\omega}$ is a subset of $(\appas)_{\ast}$.
Then,  following the reasoning below equation \eqref{eqn: pre-Hilbert product on Jomega}, if $\mathbf{P} $ is any projection such that $\omega(\mathbf{P})=0$, then  $\eta_{\mathbf{P}}=\mathbf{0}$  and thus $\eta(\mathbf{P})=l_{\eta}(\mathbf{P})=0$ which means that $\eta\in AC_{\omega}$  because of point \ref{point: AComega for W*}.

Since the norm determined by the inner product on $J_{\omega}$ is stronger than the norm on $J_{\omega}$ inherited from $(\appas)_{\ast}$, the identification map $\mathrm {i}\colon J_{\omega}\ra AC_{\omega}$ can be extended to a bounded linear map $\mathrm {i}\colon \mathcal{J}_{\omega}\ra AC_{\omega}$ whose image is clearly dense.
Therefore, the dual map $\mathrm {i}^{\ast}\colon (AC_{\omega})^{\ast}\cong \appas^{\omega}\ra (\mathcal{J}_{\omega})^{\ast}$  \cfr{ proposition \ref{prop: the dual of the tangent double cone}} is injective because
\be\label{eqn: dual of canonical immersion of Jomega in AComega is injective with dense range}
\left(\mathrm{i}^{\ast}(\mathbf{a}-\mathbf{b})\right)(\eta)=\eta(\mathbf{a}-\mathbf{b}) = 0 \;\;\forall \eta\in \mathcal{J}_{\omega}
\ee
implies that $\left(\mathrm{i}^{\ast}(\mathbf{a}-\mathbf{b})\right)(\eta_{\mathbf{a}-\mathbf{b}})=\omega((\mathbf{a}-\mathbf{b})^{2})=0$ , and, in turn, this is equivalent to $(\mathbf{a}-\mathbf{b})$ lying in the Gel'fand ideal of $\omega $ \cfr{equation \eqref{eqn: Gelfand ideal}}.
Therefore, being $\omega$ a \nplf\ and being $(\mathbf{a}-\mathbf{b})$ self-adjoint, equation \eqref{eqn: Gelfand ideal nplf} immediately implies that $(\mathbf{a}-\mathbf{b})\in (\appa_{qq})_{sa}$ so that $(\mathbf{a}-\mathbf{b})=\mathbf{0}$ because $\mathbf{a},\mathbf{b}\in\appas^{\omega}$, and thus $\mathrm {i}^{\ast}$ is injective.
Recalling that $\mathcal{J}_{\omega}$ is isomorphic to its dual beacause it is a Hilbert space, equation \eqref{eqn: dual of canonical immersion of Jomega in AComega is injective with dense range} also implies that $\mathrm {i}^{\ast}$ has a dense range so that  $(AC_{\omega})^{\ast}\cong \appas^{\omega}$   can be identified with a dense subset of the dual of $\mathcal{J}_{\omega}$ as claimed.

\qed

Once we have the inner product $G_{\omega}$ on  $\mathcal{J}_{\omega}\subseteq AC_{\omega}$ for every \nplf\ $\omega$, we are ready to add some spice to the notion of parametric model introduced in definition \ref{defn: parametric models} by requiring additional regularity conditions that allows the definition of a (possibly degenerate) Riemannian metric tensor on the manifold of the model capturing the essense of the inner product $G_{\omega}$ for each $\omega$ in the model.

\begin{definition}\label{defn: Jordan-regular parametric model}
A smooth parametric model   $( \mathcal{M}, j, \mathscr{A})$ \cfr{definition \ref{defn: parametric models}} is said to be \textbf{J-regular}  if
\be\label{eqn: Jordan metric tensor on parametric models}
\left(\mathcal{G}(X,Y)\right)(m):=G_{\mathrm{j}(m)}\left(T_{m}\mathrm{j}(X(m)),T_{m}\mathrm{j}(Y(m))\right)  
\ee
with $X,Y$ smooth vector fields on $\mathcal{M}$ and where the right-hand-side is the inner product in $ \mathcal{J}_{j(m)}$ \cfr{definition \eqref{defn: Jordan Hilbert space}}, defines a smooth tensor field on $\mathcal{M}$.
Obviously, a \textbf{J-regular} parametric model  $( \mathcal{M}, j, \mathscr{A} )$ is such that the image $T_{m}\mathrm{j}(T_{m}M)$ of the tangent space at $m\in\mathcal{M}$ throught the tangent map of j  is a subset of the real Hilbert space $  \mathcal{J}_{j(m)}$ \cfr{definition \eqref{defn: Jordan Hilbert space}} for all $m\in\mathcal{M}$.
\end{definition}

We close this section by showing that a  monotonicity property holds for $ G_{\omega}$ under norm-continuous, ultra-weakly continuous, completely-positive, unital maps. 

\begin{proposition}\label{prop: monotonicity property}
Let $\Phi\colon\appa\ra\bappa$ be a norm-continuous, ultra-weakly continuous, completely-positive, unital map. 
Let $\rho\in\pos(\appa)$ be such that $\rho=\Phi^{\ast}(\omega)$ with $\omega\in\pos(\bappa)$ and $\Phi^{\ast}\colon\bappa^{\ast}\ra\appa^{\ast}$ the dual map of $\Phi$.
For every $\eta\in\mathcal{J}_{\omega}$ it holds $\Phi^{\ast}(\eta)\in\mathcal{J}_{\rho}$ and 
\be\label{eqn: monotonicity of Jordan inner product} 
G_{\rho}(\Phi^{\ast}\eta, \Phi^{\ast}\eta)\leq G_{\omega}(\eta,\eta).
\ee
Equation \eqref{eqn: monotonicity of Jordan inner product} is referred to as the \grit{monotonicity property} of $G$ with respect to (norm-continuous and ultra-weakly continuous) completely-positive, unital maps.

\end{proposition}
\proof
First of all, we note that dual map $\Phi^{\ast}$ sends positive linear functionals into positive linear functionals because $\Phi$ is positive.
Moreover, since $\Phi$ is also ultra-weakly continuous, there is a bounded linear map $\varphi\colon \mathcal{B}(\bappa)\ra\mathcal{B}(\appa)$ between the predual of $\bappa$ and the predual of $\appa$ of which $\Phi$ is the dual map.
Therefore,    $\Phi^{\ast}$ sends normal linear functionals into normal linear functionals.
In general, if $\rho=\Phi^{\ast}(\omega)$,  the map $\Phi$ does not send $\appas^{\rho}$ into $\bappa_{sa}^{\omega}$ \cfr{proposition \ref{prop: the dual of the tangent double cone}} .
However, denoting with $\mathbf{P}_{\omega}$ the projection onto $\bappa_{sa}^{\omega}$ (which exists because $\bappa_{sa}^{\omega}$ is complemented in $\bappa_{sa}$ since $\omega$ is a \nplf), it is clear that the map
\be
\mathbf{a}\mapsto \Phi_{ \omega}(\mathbf{a}):=\mathbf{P}_{\omega}(\Phi(\mathbf{a}))
\ee
is a linear map between $\appas^{\rho}$ and $\bappa_{sa}^{\omega}$.
What is interesting is that, when looking at $\appas^{\rho}$ as a dense subspace of $\mathcal{J}_{\rho}^{\ast}$ and at $\bappa_{sa}^{\omega}$ as a dense subspace of $\mathcal{J}_{\omega}^{\ast}$ according to proposition  \ref{prop: properties of Jordan Hilbert space and its dual}, it defines a bounded linear map between $\mathcal{J}_{\rho}^{\ast}$ and $\mathcal{J}_{\omega}^{\ast}$.
Indeed, for every $\mathbf{a}\in\appas^{\rho}$, it holds
\be
||\Phi_{ \omega}(\mathbf{a})||^{2}_{\omega}=G_{\omega}(\mathbf{P}_{\omega}(\Phi(\mathbf{a})),\mathbf{P}_{\omega}(\Phi(\mathbf{a})))=\omega(\mathbf{P}_{\omega}(\Phi(\mathbf{a}))\,\mathbf{P}_{\omega}(\Phi(\mathbf{a}))).
\ee
Let us denote by $\mathbf{Q}_{\omega}$ the complement projection of $\mathbf{P}_{\omega}$.
Obviously, $\mathbf{P}_{\omega}\mathbf{Q}_{\omega}=\mathbf{Q}_{\omega}\mathbf{P}_{\omega}=\mathbf{0}$ and thus $\omega(\mathbf{a}\mathbf{b})=\omega(\mathbf{P}_{\omega}(\mathbf{a})\mathbf{P}_{\omega}(\mathbf{b})) + \omega(\mathbf{Q}_{\omega}(\mathbf{a})\mathbf{Q}_{\omega}(\mathbf{b}))$.
Moreover, since  $\mathbf{Q}_{\omega}(\bappa )=  \bappa_{qq}  $ is a \calg\ of $\bappa$  and  $\omega$ vanishes on $\bappa_{qq}$ \cfr{discussion below equation \eqref{eqn: nplf and its support projection}}, we conclude that $\omega(\mathbf{a}\mathbf{b})=\omega(\mathbf{P}_{\omega}(\mathbf{a})\mathbf{P}_{\omega}(\mathbf{b})) $ so that 
\be
||\Phi_{ \omega}(\mathbf{a})||^{2}_{\omega} = \omega(\Phi(\mathbf{a})\Phi(\mathbf{a})).
\ee
Since $\Phi$ is completely positive, it follows that $\Phi(\mathbf{a})\Phi(\mathbf{a})\leq\Phi(\mathbf{a}^{2})$ \cite{Choi-1974} and thus
\be\label{eqn: Phirhoomega is bounded}
||\Phi_{ \omega}(\mathbf{a})||^{2}_{\omega}\leq \omega(\Phi(\mathbf{a}^{2}))=\rho(\mathbf{a}^{2})=G_{\rho}(\mathbf{a},\mathbf{a})=||\mathbf{a}||_{\rho}^{2}.
\ee
Equation \eqref{eqn: Phirhoomega is bounded} implies that $\Phi_{ \omega}$ can be extended to a bounded linear map between $\mathcal{J}_{\rho}^{\ast}$ and $\mathcal{J}_{\omega}^{\ast}$ as claimed.
Moreover,  equation \eqref{eqn: Phirhoomega is bounded}  also implies that $||\Phi_{\omega}||\leq 1$ so that $\Phi_{\omega}$ is a contraction.

Now, consider the dual map  $\Phi_{ \omega}^{\ast}\colon \mathcal{J}_{\omega}\ra \mathcal{J}_{\rho}$.
Denoting with $(|)_{\bullet}$ the pairing between  $\mathcal{J}_{\bullet}$ and its dual, it holds
\be
(\mathbf{a}| \Phi_{ \omega}^{\ast}(\eta))_{\rho}=(\Phi_{ \omega}(\mathbf{a})|\eta)_{\rho}= \eta(\Phi_{ \omega}(\mathbf{a}))=\eta(\mathbf{P}_{\omega}(\Phi(\mathbf{a})))
\ee
for all $\mathbf{a}\in\appas^{\rho}\subseteq \mathcal{J}_{\rho}^{\ast}$.
According to proposition \ref{prop: properties of Jordan Hilbert space and its dual}, every $\eta\in\mathcal{J}_{\omega}$ is also an element of $AC_{\omega}$, and  proposition \ref{prop: the dual of the tangent double cone} implies that  $\eta(\Phi(\mathbf{a}))=\eta(\mathbf{P}_{\omega}(\Phi(\mathbf{a})))+ \eta(\mathbf{Q}_{\omega}(\Phi(\mathbf{a})))=\eta(\mathbf{P}_{\omega}(\Phi(\mathbf{a})))$.
Accordingly, it follows that 
\be
(\mathbf{a}| \Phi_{ \omega}^{\ast}(\eta))_{\rho}
=\eta( \Phi(\mathbf{a}))=(\Phi^{\ast}(\eta))(\mathbf{a})
\ee
which means that $\Phi^{\ast}_{\omega}$ coincides with $\Phi^{\ast}$ on $\mathcal{J}_{\omega}\subseteq AC_{\omega}$.
Eventually, recalling that $\Phi_{\omega}$ is a contraction and that the norm of $\Phi^{\ast}_{\omega}$ coincides with that of $\Phi_{\omega}$, it immediately follows that
\be 
G_{\rho}(\Phi^{\ast}(\eta), \Phi^{\ast}(\eta))=G_{\rho}(\Phi^{\ast}_{\omega}(\eta), \Phi^{\ast}_{\omega}(\eta))\leq ||\Phi_{\omega}||\,\, G_{\omega}(\eta,\eta)\leq G_{\omega}(\eta,\eta)
\ee
as claimed.
\qed

\begin{remark}\label{rem: unitary invariance}
When $\Phi$ is an ultra-weakly continuous   automorphism of $\appa$, then it is also an automorphism of the Jordan product.
Therefore, recalling that $\rho=\Phi^{\ast}(\omega)$ and that $\Phi^{\ast}_{\omega}$ coincides with $\Phi^{\ast}$ on $\mathcal{J}_{\omega}\subseteq AC_{\omega}$ \cfr{proposition \ref{prop: monotonicity property}}, a direct computation shows that
\be
\begin{split}
\left(\Phi^{\ast} (\eta_{\mathbf{a}})\right)(\mathbf{b})&= \eta_{\mathbf{a}}\left(\Phi(\mathbf{b})\right)=\omega\left(\{\mathbf{a},\Phi(\mathbf{b})\}\right)=   \omega\left(\Phi\left(\Phi^{-1}\left(\{\mathbf{a},\Phi(\mathbf{b}\}\right)\right)\right)=\\
&=\omega\left(\Phi\left(\{\Phi^{-1}\left(\mathbf{a}\right), \mathbf{b}\} \right)\right)=\rho(\{\Phi^{-1}\left(\mathbf{a}\right),\mathbf{b}\})
\end{split}
\ee 
for all $\mathbf{a},\mathbf{b}\in\appas$.
It then follows that
\be
G_{\rho}(\Phi^{\ast}(\eta_{\mathbf{a}}), \Phi^{\ast}(\eta_{\mathbf{a}}))=\rho \left(\Phi^{-1}\left(\mathbf{a}^{2}\right)\right)=\omega(\mathbf{a}^{2} )=G_{\omega}( \eta_{\mathbf{a}},\eta_{\mathbf{a}}),
\ee
which reflects the invariance under ultra-weakly continuous   automorphisms of the inner products $G_{\omega}$ with $\omega\in\pos$.
\end{remark}



\section{Examples}\label{sec: examples}

In this section we  will investigate some meaningful examples  that we believe clearly show the unifying perspective of our formalism when dealing with classical and quantum information geometry.

\subsection{The finite-dimensional case}\label{sec:Riemannian_Geometry_of_parametrized_models}

We briefly discuss the peculiarities of the finite-dimensional case, referring to  \cite{C-J-S-2020-02,C-J-S-2020,C-J-S-2022} for the proofs of all the results mentioned.
When $\appa$ is finite-dimensional, there is a sort of “preferred” family of Jordan regular parametric models naturally emerging when we investigate the Jordan-analogue of Konstant-Kirillov-Souriau theory of coadjoint orbits of a Lie group \cite{Kirillov-1962,Kirillov-1976,Kostant-1970,Souriau-1970}.
Specifically, since $\appa$ is finite-dimensional, all linear functional are normals so that $\appa^{\ast}=\appa_{\ast}$ and $\appas^{\ast}=(\appas)_{\ast}$.
Therefore,  in complete analogy with the Lie-algebra case, the algebraic structure of Jordan algebra on $\appas$  translates into a geometric structure on $\appas^{\ast}$, namely, a  $(2,0)$-contravariant tensor $\mathcal{R}$ whose value at $(\mathbf{a},\mathbf{b})\in T_{\xi}^{\ast}\appas^{\ast}\times  T_{\xi}^{\ast}\appas^{\ast}\cong \appas\times\appas$ is given by 
\be\label{eqn: Jordan tensor}
\mathcal{R}_{\xi}(\mathbf{a},\mathbf{b}):=\xi\left(\{\mathbf{a},\mathbf{b}\}\right).
\ee
The contravariant tensor $\mathcal{R}$ is evidently smooth (because linear in $\xi$) and symmetric because it's defined in terms of the symmetric Jordan product, and it is referred to, quite obviously, as the Jordan tensor on $\appas^{\ast}$.
The Jordan tensor $\mathcal{R}$ allows to define a generalized distribution  $\mathcal{D}=\{\mathcal{D}_{\xi}\}_{\xi\in\appas^{\ast}}$ on $\appas^{\ast}$ according to
\be\label{eqn: Jordan distribution}
\mathcal{D}_{\xi}:=\{\eta\in T_{\xi} \appas^{\ast}\cong  \appas^{\ast}\,|\;\;\exists\,\mathbf{a}\in\appas\;\colon\eta(\mathbf{b})=\mathcal{R}_{\xi}(\mathbf{a},\mathbf{b})\;\forall\,\mathbf{b}\in\appas\},
\ee
and, when instead of a generic $\xi\in\appas^{\ast}$ we consider a \nplf\ $\omega$,  equation \eqref{eqn: Jordan distribution} and equation \eqref{eqn: Jomega} immediately implies that $\mathcal{D}_{\omega}=J_{\omega}=AC_{\omega}$, where the last equality follows from the fact that  $J_{\omega}$ is closed when $\appa$ is finite-dimensional.
The distribution  $\mathcal{D}=\{\mathcal{D}_{\xi}\}_{\xi\in\appas^{\ast}}$ is referred to as the canonical distribution generated by the tensor $\mathcal{R} $.
 
 
According to \cite{C-J-S-2022}, in analogy with the Konstant-Kirillov-Souriau theory of coadjoint orbits of a Lie group, it is possible to find maximal leaves of the canonical distribution on which the tensor $\mathcal{R}$ is invertible, and its inverse is a symmetric analogue  of the Konstant-Kirillov-Souriau symplectic form on coadjoint orbits.
However, the canonical distributions of $\mathcal{R}$ does not generate a foliation of $ \appas^{\ast}$ as instead happens in the case of the symplectic foliation associated with the Poisson tensor on the dual of a Lie algebra in the Konstant-Kirillov-Souriau theory.
The tangent space at each point of any of the maximal leaves of $\mathcal{D}=\{\mathcal{D}_{\xi}\}_{\xi\in\appas^{\ast}}$ mentioned before is spanned by the tangent vectors as $\eta_{\mathbf{a}}$ in equation \eqref{eqn: elements in Jomega}, and the pointwise inverse of $\mathcal{R}$ is as by equation \eqref{eqn: FRBH inner product}.
The whole space of $\pos$ \nplfs\ of $\appa$ is decomposed into the disjoint union of such maximal leaves, and the inverse of $\mathcal{R}$ on any such leave is a Riemannian metric tensor which coincides with the  Fisher-Rao metric tensor on positive measures when $\appa$ is commutative, and with the Bures-Helstrom metric tensor on faithful (non-normalized) quantum states when $\appa=\bh$ for some finite-dimensional Hilbert space $\hh$ (a fact also noted in \cite{C-J-S-2020-02,C-J-S-2020}).

The family of maximal leaves of $\mathcal{D}=\{\mathcal{D}_{\xi}\}_{\xi\in\appas^{\ast}}$  providing a decomposition of $\pos$ is precisely the “preferred” family of Jordan regular parametric models alluded to in the beginning of this section.
It is also worth mentioning that, by suitably adapting the analysis carried out in \cite{DA-F-2021}, it could be possible to show that this preferred family provides a Withney stratification of $\pos$.

In the infinite dimensional case, quite unsurprisingly, things are more complicated  and this picture breaks down.
That is why, in this work, we proposed a change of perspective to tackle the infinite-dimensional case.
Specifically, instead of looking to generalize the foliation picture described before, we decided to simply look at the  maximal leaves of \nplfs\ of the canonical distribution of $\mathcal{R}$ in finite dimensions  as parametric models in the sense of definition \ref{defn: parametric models} endowed with a Riemannian metric tensor determined, essentially, by the Jordan product in $\appas$ through the inverse of the tensor $\mathscr{R}$.
Then, this reformulation of the finite-dimensional case led us to the notion of Jordan regular parametric model given in definition \ref{defn: Jordan-regular parametric model},  which works also  in the infinite-dimensional case.

\begin{remark}
Concerning the infinite-dimensional generalization of the foliation picture, a   relevant thing to note is that, by focusing on $(\appas)_{\ast}$, the algebraic structure of Jordan algebra of $\appas$ determines a $(2,0)$-contravariant tensor $\mathcal{R}$  on $(\appas)_{\ast}$ again by equation \eqref{eqn: Jordan tensor} essentially because $T_{\xi}^{\ast}(\appas)_{\ast}\cong\appas$.
Moreover, the map $\sharp\colon \appas\cong  T_{\xi}^{\ast}(\appas)_{\ast}\ra T_{\xi}^{\ast\ast}(\appas)_{\ast} \cong\appas^{\ast}$ given by
\be
\left(\sharp(\mathbf{a})\right)(\mathbf{b}):=\mathcal{R}_{\xi}(\mathbf{a},\mathbf{b})=\xi(\{\mathbf{a},\mathbf{b}\})
\ee
is such that the image of $T_{\xi}^{\ast}(\appas)_{\ast}$ is actually contained in $(\appas)_{\ast}\subseteq\appas^{\ast}\cong T_{\xi}^{\ast\ast}(\appas)_{\ast}$.
Therefore, $\mathcal{R}$ satisfies a condition analogous to the one exploited in \cite{B-R-2005} to define a Banach Lie-Poisson structure on $\appas$ in terms of the antysimmetric tensor built out of the Lie algebra structure on $\appas$ associated with the (scaled) commutator product in $\appa$.
However, a rigorous analysis of the geometry of the distribution generated by $\mathcal{R}$ would reasonably require particular care for technical details because, among other things, the vector space $\mathcal{D}_{\xi}$ is in general not closed.
We believe it would be very interesting to try to understand how far we can push the finite-dimensional theory developed in \cite{C-J-S-2022} into the infinite-dimensional realm to see how many “preferred” Jordan regular parametric models are obtained.
We hope we will be able to investigate this issue in the future, and we also hope our current work will inspire someone else to investigate these matters.
\end{remark}

\subsection{Classical statistical models}

The aim of this subsection is to briefly check that the formalism developed in this work is a genuine generalization of the formalism of classical information geometry as rigorously formalized in \cite{A-J-L-S-2015,A-J-L-S-2017,A-J-L-S-2018}.
Let  $\appa=\mathcal{L}^{\infty}(\mathcal{X},\mu)$ be the Abelian \walg\ recalled in section \ref{sec: walgs and nplfs}.
Recall that  normal linear functionals are identified with complex measures through equation \eqref{eqn: normal linear functionals as complex measures in the Abelian case}, and   \nplfs\ can be identified with the space of non-negative functions in $L^{1}(\mathcal{X},\mu)$.

Given a \nplf\ $\omega$ on $\appa$, the tangent double cone $T_{\omega}\pos$ is closed and it can be identified with $L^{1}(\mathcal{X},\mu_{\omega})$ \cfr{remark \ref{rem: classical tangent double cone is closed}}, that is, every $\xi\in T_{\omega}\pos$ acts as
\be
\xi(f)=\int_{\mathcal{X}}f\,\mathrm{d}\mu_{\xi}=\int_{\mathcal{X}}f\,F_{\xi}\,\mathrm{d}\mu_{\omega}
\ee
with $F_{\xi}\in L^{1}(\mathcal{X},\mu_{\omega})$ the Radon-Nikodym derivative of $\mu_{\xi}$ with respect to $\mu_{\omega}$.
If $(\mathcal{M}, \mathbf{ p },\appa)$ is a parametric model according to definition \ref{defn: parametric models}, then $\mathbf{ p }(m)=p_{m}\mu$ with $p_{m}$ a positive function in $L^{1}(\mathcal{X},\mu)$, and this, in turn, means that $(\mathcal{M}, \mathcal{X}, \mu, \mathbf{p})$ is a \emph{parametrized measure model dominated by $\mu$}   in the languange of classical information geometry \cite[defn. 3.4, p. 150]{A-J-L-S-2017}.
In particular, every $T_{m}\mathbf{ p }(v_{m})\in T_{\mathbf{ p }(m)}\pos$  leads to a Radon-Nikodym derivative  $\frac{\mathrm{d}T_{m}\mathbf{ p }(v_{m})}{\mathrm{d}\mathbf{p}(m)}$ which is nothing but the logarithmic derivative of $\mathbf{p}$ at $m$  in the direction $v_{m}$ as defined in \cite[defn. 3.6, p. 152]{A-J-L-S-2017}. 

On the other hand, it is clear that every characteristic function $1_{A}$ associated with the measurable subset  $A\subseteq \mathcal{X}$ with   $\mu_{\omega}(A)<\infty$ is in $\appa=\mathcal{L}^{\infty}(\mathcal{X},\mu)$ and determines an element $\xi_{A}\in J_{\omega}$ through 
\be
\xi_{A}(f)= \omega(\{1_{A},f\})=\int_{A}f\, \mathrm{d}\mu_{\omega}.
\ee
Accordingly,  the Hilbert space $\mathcal{J}_{\omega}$ introduced in definition \ref{defn: Jordan Hilbert space}  can be identified with the real Hilbert space $L^{2}_{\mathbb{R}}(\mathcal{X},\mu_{\omega})$.
Note that this is consistent with the well-known fact that $L^{q}(\mathcal{X},\mu_{\omega})\subseteq L^{p}(\mathcal{X},\mu_{\omega})$ for all $1\leq p \leq q \leq \infty $.
It then follows that, if the parametric model $(\mathcal{M}, \mathbf{ p },\appa)$ is  Jordan regular in the sense of definition \ref{defn: Jordan-regular parametric model}, then $(\mathcal{M}, \mathcal{X}, \mu, \mathbf{p})$ is 2-integrable in the language of  \cite{A-J-L-S-2017} and  the tensor $\mathcal{G}$ defined as in equation \ref{eqn: Jordan metric tensor on parametric models} becomes
\be
\left(\mathcal{G}(X,Y)\right)(m)=\int_{\mathcal{X}}\,\frac{\mathrm{d}T_{m}\mathbf{ p }(X_{m})}{\mathrm{d}\mathbf{p}(m)}\,\frac{\mathrm{d}T_{m}\mathbf{ p }(Y_{m})}{\mathrm{d}\mathbf{p}(m)}\,\mathrm{d}\mathbf{p}(m)
\ee
which coincides with the Fisher-Rao metric tensor as given in \cite[eqn. 3.41, p. 136]{A-J-L-S-2017}.

\subsection{Rank-one, strongly-continuous unitary models}


Consider the \walg\ $\appa=\bh$ of bounded linear operator on the possibly infinite-dimensional, separable complex Hilbert space $\hh$.
As recalled in section \ref{sec: walgs and nplfs}, $\appa=\bh$ is the Banach dual of the space $\mathcal{T}(\hh)$ of trace-class operators on $\hh$.
Therefore, every non-negative trace-class operator $\omega$ defines a \nplf\ on $\appa=\bh$ through $\omega(\mathbf{a})=\mathrm{Tr}(\omega\mathbf{a})$.

In this section, we will discuss a particular type of parametric models stemming from strongly continuous unitary representations of Banach-Lie groups.
This family of parametric models is general enough to encompass most of the parametric models used in the literature when dealing with a quantum system described in terms of a separable Hilbert space \cite{Fujiwara-1999,Holevo-2001,Holevo-2011}.

Let $\mathrm{G}$ be a Banach-Lie group and $\pi\colon \mathrm{G}\ra\mathscr{U}(\hh)$ a strongly-continuous unitary representation of $\mathrm{G}$ on $\hh$, that is, a group homomorphism between $ \mathrm{G}$ and the unitary group $\mathscr{U}(\hh)$ such that the map $\pi_{\psi}\colon\mathrm{G}\ra \hh$ given by $\gr\mapsto (\pi(\gr))(\psi)$ is continuous with respect to the norm topology in $\hh$ for every $\psi\in\hh$.
A vector $\varphi\in\hh$ is called \grit{smooth} for $\pi$ if the map $\gr\mapsto (\pi(\gr))(\varphi)$ is smooth.
In particular, if $\mathrm{G}$ is a finite-dimensional Lie group, then smooth vectors always exist and are dense in $\hh$ \cite{Garding-1947}.

\begin{proposition}\label{prop: rank-one strongly-continuous unitary jordan-regular models}
If $\varphi$ is a smooth vector for the strongly-continuous unitary representation $\pi\colon \mathrm{G}\ra\mathscr{U}(\hh)$, then $(\mathrm{G},\jj,\appa=\bh )$, with $\jj(\gr):=|(\pi(\gr))(\varphi)\rangle\langle (\pi(\gr))(\varphi)| \equiv | \varphi_{\gr}\rangle\langle  \varphi_{\gr}| $ is a J-regular parametric model according to definition \ref{defn: Jordan-regular parametric model}.
Moreover,  the tensor $\mathcal{G}$ defined as in equation \eqref{eqn: Jordan metric tensor on parametric models} is invariant with respect to the canonical left action of $\mathrm{G}$ on itself.

\end{proposition}

\proof
We first prove that $(\mathrm{G},\jj,\bh )$ is a smooth parametric model according to definition \ref{defn: parametric models}.
It suffices to  note that $\jj$ can be written as the composition between the smooth map $\pi_{\varphi}\colon \mathrm{G}\ra\hh$  given by $\pi_{\varphi}(\gr):=(\pi(\gr))(\varphi)$, and the smooth map from $\hh$ to the space $\mathcal{T}_{sa}(\hh)$ of self-adjoint trace-class operators on $\hh$ given by $\psi\ra F(\psi)=|\psi\rangle\langle\psi|$.

Then,  it holds
\be
T_{\gr}\pi_{\varphi}(v_{\gr})=\frac{\mathrm{d}}{\mathrm{d}t}\left((\pi(\gr_{t}))(\varphi)\right)_{t=0}
\ee
for an arbitrary tangent vector  $v_{\gr}\in T_{\gr}\mathrm{G}$, and it holds
\be
T_{\psi}F (\phi) = \frac{\mathrm{d}}{\mathrm{d}t}\left(|\psi_{t}\rangle\langle\psi_{t}|\right)_{t=0}=  |\phi\rangle\langle\psi| +|\psi\rangle\langle\phi| 
\ee
for an arbitrary tangent vector  $\phi\in T_{\psi}\hh\cong\hh$, so that
\be\label{eqn: tangent vector rank-1 unitary models}
T_{\gr}\jj(v_{\gr})=T_{\pi_{\varphi}(\gr)}F \circ T_{\gr}\pi_{\varphi}(v_{\gr})=| v_{\gr} \rangle\langle\varphi_{\gr}| + |\varphi_{\gr}\rangle\langle  v_{\gr} |
\ee
where, with an evident abuse of notation, on the right-hand side, we denoted by $T_{\gr}\pi_{\varphi}(v_{\gr})$ simply as $v_{\gr}$.
Since $\langle \varphi_{\gr}|\varphi_{\gr}\rangle=C_{\varphi}$ is a constant depending only on the reference vector $\varphi$ and not  on $\gr$, it also holds 
\be\label{eqn: infinitesimal normalization condition rank-1 unitary models}
\langle v_{\gr} |\varphi_{\gr}\rangle + \langle\varphi_{\gr}|  v_{\gr }\rangle = 0
\ee
for every $\gr\in\mathrm{G}$ and every $v_{\gr}\in T_{\gr}\mathrm{G}$.

Being $\mathrm{G}$ a Banach-Lie group, there is a natural left action $L$  of $\mathrm{G}$ on itself, and it follows that
\be
\jj\circ L_{\mathrm{h}} = Ad_{\pi(\mathrm{h})} \circ\jj
\ee
with $Ad_{\pi(\mathrm{h})} (\mathbf{a})=\pi(\mathrm{h})\mathbf{a}\pi(\mathrm{h})^{\dagger}$ for every $\mathbf{a}\in\appa$.
Consequently, it also follows that
\be\label{eqn: infinitesimal version of equivariance in rank1 unitary models}
T_{\gr}\left(\jj\circ L_{\mathrm{h}} \right)= T_{\gr}\left( Ad_{\pi(\mathrm{h})} \circ\jj\right)=T_{\jj(\gr)}  Ad_{\pi(\mathrm{h})} \circ T_{\gr}\jj = Ad_{\pi(\mathrm{h})} \circ T_{\gr}\jj
\ee
where the last equality basically follows from the fact that $ Ad_{\pi(\mathrm{h})}  $ is linear.

Again because $\mathrm{G}$ is a Banach-Lie group, in order to write down $\mathcal{G}$ following equation \eqref{eqn: Jordan metric tensor on parametric models}, we can focus only on left-invariant vector fields.
If $X$ is a left-invariant vector field on $\mathrm{G}$, it holds
\be
X(\gr)= T_{\mathrm{e}}L_{\gr} (X(\mathrm{e}))
\ee
where $\mathrm{e}$ is the identity element in $\mathrm{G}$.
Accordingly, exploiting equation \eqref{eqn: infinitesimal version of equivariance in rank1 unitary models} we obtain
\be
T_{\gr}\jj (X(\gr))=  T_{\gr}\jj\circ T_{\mathrm{e}}L_{\gr} (X(\mathrm{e}))= \left(T_{\gr}(\jj\circ L_{\gr})\right)(X(\mathrm{e}))=Ad_{\pi(\mathrm{g})} \circ T_{\gr}\jj(X(\mathrm{e})).
\ee
Now, we note that $Ad_{\pi(\mathrm{g})}$ can be thought of as the dual of a norm-continuous, ultra-weakly continuous automorphism of $\appa=\bh$, and thus, according to remark \ref{rem: unitary invariance}, it holds
\be\label{eqn: invariance of G for rank1 unitary models}
G_{\jj(\gr)}(T_{\gr}\jj(X(\gr)),T_{\gr}\jj(Y(\gr)))=
G_{\jj(\mathrm{e})}(T_{\mathrm{e}}\jj(X(\mathrm{e})),T_{\mathrm{e}}\jj(Y(\mathrm{e})))
\ee
for all left-invariant vector fields $X,Y$ on $\mathrm{G}$.
Equation \eqref{eqn: invariance of G for rank1 unitary models} implies that we only need to understand what happens to tangent vectors at the identity element $\mathrm{e}$ in order to understand $\mathcal{G}$ \cfr{equation \eqref{eqn: Jordan metric tensor on parametric models}}.

Let $\{\phi_{j}\}_{j\in\mathbb{N}}$ be an orthonormal basis in $\hh$ such that $\varphi=A\phi_{0}$.
According to equation \eqref{eqn: tangent vector rank-1 unitary models}, for every left-invariant vector field $X$ on $\mathrm{G}$ there is a vector $v_{e}\in\hh$ such that 
\be
T_{\mathrm{e}}\jj(X(\mathrm{e}))=|\varphi\rangle\langle v_{e}| + |v_{e}\rangle\langle\varphi|,
\ee
and we write
\be
v_{e}= v_{e}^{0}\phi_{0} + \sum_{k=1}^{\infty}v_{e}^{k}\phi_{k} \equiv v_{e}^{0}\phi_{0} +v_{e}^{\perp}.
\ee
Then, define $\mathbf{X}_{\mathrm{e}}\in\appas=(\bh)_{sa}$ according to
\be
\mathbf{X}_{\mathrm{e}}=\frac{ \overline{A}v_{e}^{0} + A\overline{v_{e}^{0}}}{|A|^{2}}\,|\phi_{0}\rangle\langle\phi_{0}| + \frac{1}{2|A|^{2}}\left(|\varphi\rangle\langle v_{e}^{\perp}| + |v_{e}^{\perp}\rangle\langle\varphi|\right),
\ee 
so that, since $\jj(\mathrm{e})=|\varphi\rangle\langle\varphi|$, it is a matter of straightforward computation to check that  such that
\be
|\varphi\rangle\langle v_{e}| + |v_{e}\rangle\langle\varphi|= T_{\mathrm{e}}\jj(X(\mathrm{e}))=\{\jj(\mathrm{e}),\mathbf{X}_{\mathrm{e}}\}
\ee
from which it follows
\be\label{eqn: tangent vector as anticommutator rank-1 unitary models}
\Tr(T_{\mathrm{e}}\jj(X(\mathrm{e}))\mathbf{a}) =\Tr(\jj(\mathrm{e}) \{\mathbf{X}_{\mathrm{e}},\mathbf{a}\})
\ee
for all $\mathbf{a}\in\appa=\bh$.
At this point, recalling equation \eqref{eqn: FRBH inner product}, it follows that
\be
G_{\jj(\mathrm{e})}(T_{\mathrm{e}}\jj(X(\mathrm{e})),T_{\mathrm{e}}\jj(Y(\mathrm{e})))=\Tr(\jj(\mathrm{e}) \{\mathbf{X}_{\mathrm{e}},\mathbf{Y}_{\mathrm{e}}\}).
\ee
Replacing $X(\mathrm{e})$ with $Y(\mathrm{e})$ and $\mathbf{a}$ with $\mathbf{X}_{\mathrm{e}}$ in  equation \eqref{eqn: tangent vector as anticommutator rank-1 unitary models}, end exploiting equation \eqref{eqn: tangent vector rank-1 unitary models} for $T_{\mathrm{e}}\jj(Y(\mathrm{e}))$, it follows that
\be
G_{\jj(\mathrm{e})}(T_{\mathrm{e}}\jj(X(\mathrm{e})),T_{\mathrm{e}}\jj(Y(\mathrm{e})))=\Tr(T_{\mathrm{e}}\jj(Y(\mathrm{e}))\mathbf{X}_{\mathrm{e}}) =C_{\varphi} (\langle\varphi|\mathbf{X}_{\mathrm{e}}|              Y(\mathrm{e}) \rangle +  \langle  Y(\mathrm{e}) |\mathbf{X}_{\mathrm{e}}|\varphi \rangle)
\ee
where $\langle \varphi |\varphi \rangle=C_{\varphi}$.
Now, to get rid of the dependence from the “uknown” $\mathbf{X}_{\mathrm{e}}$ we  exploit again equation  \eqref{eqn: tangent vector as anticommutator rank-1 unitary models}, but this time replacing only $\mathbf{a}$ with  $| Y(\mathrm{e}) \rangle\langle\varphi | + |\varphi \rangle\langle  Y(\mathrm{e}) |$.
A direct computation shows that the left-hand side becomes
\be
\Tr(T_{\mathrm{e}}\jj(X(\mathrm{e}))\mathbf{a})=C_{\varphi}(\langle X(\mathrm{e}) | Y(\mathrm{e}) \rangle + \langle Y(\mathrm{e}) | X(\mathrm{e}) \rangle) + 2\langle X(\mathrm{e}) |\varphi\rangle\langle Y(\mathrm{e}) |\varphi \rangle    
\ee
where we also used equation \eqref{eqn: infinitesimal normalization condition rank-1 unitary models}, while the right-hand side becomes
\be
\Tr(\jj(\gr) \{\mathbf{X}_{\mathrm{e}},\mathbf{a}\})=\frac{C_{\varphi}}{2} \left(\langle \varphi |\mathbf{X}_{\mathrm{e}}|Y(\mathrm{e})\rangle + \langle Y(\mathrm{e})|\mathbf{X}_{\mathrm{e}}|\varphi \rangle  \right)
\ee
where we also used equation \eqref{eqn: infinitesimal normalization condition rank-1 unitary models}.
Eventually, we obtain 
\be
G_{\jj(\mathrm{e})}(T_{\mathrm{e}}\jj(X(\mathrm{e})),T_{\mathrm{e}}\jj(Y(\mathrm{e})))=2 C_{\varphi}(\langle X(\mathrm{e})| Y(\mathrm{e})\rangle + \langle Y(\mathrm{e})|X(\mathrm{e})\rangle) + 4\langle X(\mathrm{e})|\varphi \rangle\langle Y(\mathrm{e})|\varphi \rangle     .
\ee
Consequently, recalling equation equation \eqref{eqn: Jordan metric tensor on parametric models} and equation \ref{eqn: invariance of G for rank1 unitary models}, we obtain
\be
\left(\mathcal{G}(X,Y)\right)(\gr)=2 C_{\varphi}(\langle X(\mathrm{e})| Y(\mathrm{e})\rangle + \langle Y(\mathrm{e})|X(\mathrm{e})\rangle) + 4\langle X(\mathrm{e})|\varphi \rangle\langle Y(\mathrm{e})|\varphi \rangle  
\ee
which clearly shows that $\mathcal{G}$ is indeed a smooth tensor field on $\mathrm{G}$ which is invariant with respect to the canonical left action of $\mathrm{G}$ on itself. 
Eventually, we conclude that $(\mathrm{G},\jj,\bh)$ is   a J-regular parametric model according to definition \ref{defn: Jordan-regular parametric model} as claimed.
\qed

\begin{corollary}\label{prop: rank-one strongly-continuous unitary jordan-regular models 2}
Let $(G,\jj,\bh)$ be a J-regular parametric model as in proposition \ref{prop: rank-one strongly-continuous unitary jordan-regular models}.
Assume that the isotropy subgroup $G_{\varphi}$ of $\varphi\in\hh$ with respect to $\pi$ is such that $G/G_{\varphi}\cong \mathcal{M}$ is a smooth manifold  for which there exists a smooth section $\sigma\colon   \mathcal{M}\ra G$.
Then $( \mathcal{M},\jj\circ\sigma, \bh)$ is a J-regular parametric model \cfr{definition \ref{defn: Jordan-regular parametric model}} whose tensor field $mathcal{G}$ is invariant with respect to the canonical action of $G$ on $\mathcal{M}$.
\end{corollary}

When $G=\Uh$ and $\pi$ is the identity map, then proposition \ref{prop: rank-one strongly-continuous unitary jordan-regular models} gifts $\Uh$ a left invariant degenerate Riemannian metric tensor.
Moreover, corollary \ref{prop: rank-one strongly-continuous unitary jordan-regular models 2} gifts the complex projective space $\mathbb{CP}(\hh)\cong\Uh/U(1)$ a Riemannian metric tensor $\mathcal{G}$ which is invariant under the canonical action of $\Uh$ on $\mathbb{CP}(\hh)$.
This invariance property is strong enough to force $\mathcal{G}$ to be a constant multiple of the Fubini-Study metric tensor and we thus obtain the standard Riemannian structure of the space of normal pure states that is well-known in the context of geometric Quantum Mechanics \cite{A-S-1999,C-M-P-1990,E-M-M-2010,Kibble-1979}.

When $G$ is a finite-dimensional Lie group and $\pi$ a strongly continuous unitary representation for which both proposition proposition \ref{prop: rank-one strongly-continuous unitary jordan-regular models} and corollary  \ref{prop: rank-one strongly-continuous unitary jordan-regular models 2} are valid, then we obtain parametric models of \nplfs that are basically the well-known and widely used coherent states \cite{A-A-G-1999,Perelomov-1986}.

\section{Conclusions}\label{sec: conclusions}

The theory of J-regular parametric  models of \nplfs\ on a \walg\ $\appa$ introduced here is only at a preliminary stage, and very much is yet there to be investigated.
For instance, the formulation of the Cramer-Rao bound in the context of \walgs\ is definitely something that should be addressed.
Some preliminary results in the finite-dimensional case are discussed in \cite{C-J-S-2020-02}, from which it follows that an infinite-dimensional reformulation would necessarily entail the investigation of the so-called theory of quantum measurement in the context of information geometry.

Then, it is reasonable to ask what happens in the $W^{\star}$-algebraic framework to other well-known geometrical structures appearing in the classical case, first and foremost, to the Amari-Cencov tensor determining the dualistic structure in the classical case.
At this purpose, let us simply note that, when $\omega$ is faithful, the Jordan structure of $\appas$ gives rise to a trilinear map $T_{\omega}$ on $J_{\omega}$ by means of the so-called Jordan triple product
\be\label{eqn: canonical jordan triple product}
\{\mathbf{a},\mathbf{b},\mathbf{c}\}=\{\{\mathbf{a},\mathbf{b}\},\mathbf{c}\} + \{\mathbf{a},\{\mathbf{b},\mathbf{c}\}\} -\{\mathbf{b},\{\mathbf{a},\mathbf{c}\}\} =\frac{1}{2}\left(\mathbf{a}\mathbf{b}\mathbf{c} + \mathbf{b}\mathbf{c}\mathbf{a}\right) .
\ee
The Jordan triple product is trilinear, it is symmetric in the outer variables, and it satisfies
\be\label{eqn: property of jordan triple product}
\{\mathbf{a},\mathbf{b},\{\mathbf{x},\mathbf{y},\mathbf{z}\}\}=\{\{\mathbf{a},\mathbf{b},\mathbf{x}\},\mathbf{y},\mathbf{z}\} -\{\mathbf{x},\{\mathbf{b},\mathbf{a},\mathbf{y}\},\mathbf{z}\} + \{\mathbf{x},\mathbf{y},\{\mathbf{a},\mathbf{b},\mathbf{z}\}\}.
\ee
Actually, we could define a Jordan triple product on a vector space $V$ which is not a Jordan algebra (that is, does not have a Jordan product) simply taking a trilinear map which is symmetric in the outer variables and satisfies equation \eqref{eqn: property of jordan triple product}.
When $V$ is already a Jordan algebra and the Jordan triple product is defined through equation \eqref{eqn: canonical jordan triple product}, it is called \grit{canonical}.
 
The map $T_{\omega}$ is then defined as
\be
T_{\omega}(\eta_{\mathbf{a}},\eta_{\mathbf{b}},\eta_{\mathbf{c}}):=\omega\left(\{\mathbf{a},\mathbf{b},\mathbf{c}\}\right)= \frac{1}{2}\omega\left(\mathbf{a}\mathbf{b}\mathbf{c} + \mathbf{b}\mathbf{c}\mathbf{a}\right),
\ee
where $\eta_{\mathbf{a}},\eta_{\mathbf{b}},\eta_{\mathbf{c}}$ are as in equation \eqref{eqn: elements in Jomega}.
When $\omega$ is faithful, then  $\eta_{\mathbf{a}}=\mathbf{0}$ implies $\mathbf{a}=\mathbf{0}$ so that $T_{\omega}(\eta_{\mathbf{a}},\eta_{\mathbf{b}},\eta_{\mathbf{c}}) =0$ for every $\eta_{\mathbf{b}},\eta_{\mathbf{c}}\in J_{\omega}$ (and similarly if we replace $\mathbf{a}$ with $\mathbf{b}$ or with $\mathbf{c}$).
Therefore, we conclude that $T_{\omega}$ is indeed well-defined on $J_{\omega}\times J_{\omega}\times J_{\omega}$, and it is then evident from its very definition that $T_{\omega}$ is multilinear and that it is symmetric in the outer variable, just as the canonical Jordan triple product.
In the finite-dimensional, Abelian case, a direct computation using any Cartesian coordinate system on $(\appas)^{\ast}\cong\mathbb{R}^{n}$ immediately shows that $T_{\omega}$ reduces to the Amari-Cencov tensor on the manifold of strictly-positive probability distributions.
The investigation of the other case thus seems to be promising and to point to an even deeper connection between the Jordan algebra structure of $\appas$ and the geometrical structures of information geometry.

Another possible avenue of investigation is fueled by the so-called Petz's classification of monotone quantum metric tensors in the finite-dimensional case \cite{Petz-1996}.
It turns out that, in the finite-dimensional quantum case, there is more than one Riemannian metric tensor which is meaningful with respect to information-theoretic tasks.
For instance, beside the Bures-Helstrom metric tensor (obtained here through the Jordan product of $\appas$), it is worth mentioning at least   the Wigner-Yanase metric tensor \cite{G-I-2001,G-I-2003,Hasegawa-2003}, the Bogoliubov-Kubo-Mori metric tensor \cite{N-V-W-1975,Petz-1994}, and the quantum Tsallis  metric tensors \cite{M-M-V-V-2017}.
It is then reasonable to ask what happens to all these metric tensors in the $W^{\star}$-algebraic framework  introduced here,  if/how it is possible to describe them in terms of algebraic structures, and if it is possible to (even only partially) to extend Petz's classification to this framework.

Finally, we mention the possibility of investigating Cencov's beautiful categorical approach to information theory, statistics, and inference \cite{Cencov-1982} in the unifying framework of \walgs\ by developing a suitable category that is able to deal with the classical and quantum case simultaneously.
This point is perhaps the most ambitious one, but it also comes with unexpected connection with very recent categorical investigations on classical and quantum information theory and statistics   \cite{Fritz-2020,F-G-P-2021,F-G-P-R-2020,Parzygnat-2020} and with a recently developed formalism of quantum mechanics based on groupoids and their algebras \cite{C-DC-I-M-2020-02,C-I-M-2018,C-I-M-2019-02}.


\section*{Funding}

F. M. C. acknowledges that this work has been supported by the Madrid Government (Comunidad de Madrid-Spain) under the Multiannual Agreement with UC3M in the line of “Research Funds for Beatriz Galindo Fellowships” (C\&QIG-BG-CM-UC3M), and in the context of the V PRICIT (Regional Programme of Research and Technological Innovation).
He also wants to thank the incredible support of the Max Planck Institute for the Mathematics in the Sciences in Leipzig where he was formerly employed when this work was initially started and developed.
L. S. acknowledges partial support by grant SCHW893/5-1 of the
Deutsche Forschungsgemeinschaft, and also expresses his gratitude for the hospitality of the Max Planck Institute for the Mathematics in the Sciences in Leipzig during numerous visits.


\addcontentsline{toc}{section}{References}
{\footnotesize
\begin{thebibliography}{10}

\bibitem{A-M-R-1988}
R.~Abraham, J.~E. Marsden, and T.~Ratiu.
\newblock {\em {Manifolds, tensor analysis, and applications}}.
\newblock Springer-Verlag, New York, second edition, 1988.
\newblock {\href{https://doi.org/10.1007/978-1-4612-1029-0}{DOI:
  10.1007/978-1-4612-1029-0}}.

\bibitem{A-S-2001}
E.~M. Alfsen and F.~W. Shultz.
\newblock {\em {State spaces of operator algebras}}.
\newblock Springer-Verlag, New York, 2001.
\newblock {\href{https://doi.org/10.1007/978-1-4612-0147-2}{DOI:
  10.1007/978-1-4612-0147-2}}.

\bibitem{A-A-G-1999}
T.~S. Ali, J.P. Antoine, and J.P. Gazeau.
\newblock {\em {Coherent states, wavelets, and their geeralizations}}.
\newblock Springer-Verlag, New York, 1999.
\newblock {\href{https://doi.org/10.1007/978-1-4614-8535-3}{DOI:
  10.1007/978-1-4614-8535-3}}.

\bibitem{Amari-2016}
S.~I. Amari.
\newblock {\em {Information Geometry and its Application}}.
\newblock Springer, Japan, 2016.
\newblock {\href{https://doi.org/10.1007/978-4-431-55978-8}{DOI:
  10.1007/978-4-431-55978-8}}.

\bibitem{A-N-2000}
S.~I. Amari and H.~Nagaoka.
\newblock {\em {Methods of Information Geometry}}.
\newblock American Mathematical Society, Providence, RI, 2000.
\newblock {\href{https://doi.org/10.1090/mmono/191}{DOI: 10.1090/mmono/191}}.

\bibitem{A-S-1999}
A.~Ashtekar and T.~A. Schilling.
\newblock Geometrical formulation of quantum mechanics.
\newblock In A.~Harvey, editor, {\em On Einstein's Path: Essays in Honor of
  Engelbert Schucking}, pages 23 -- 65. Springer-Verlag, New York, 1999.

\bibitem{A-J-L-S-2015}
N.~Ay, J.~Jost, H.~V. Le, and L.~Schwachh\"{o}fer.
\newblock {Information geometry and sufficient statistics}.
\newblock {\em Probability Theory and Related Fields}, 162(1):327-- 364, 2015.

\bibitem{A-J-L-S-2017}
N.~Ay, J.~Jost, H.~V. Le, and L.~Schwachh\"{o}fer.
\newblock {\em {Information Geometry}}.
\newblock Springer International Publishing, 2017.
\newblock {\href{https://doi.org/10.1007/978-3-319-56478-4}{DOI:
  10.1007/978-3-319-56478-4}}.

\bibitem{A-J-L-S-2018}
N.~Ay, J.~Jost, H.~V. Le, and L.~Schwachh\"{o}fer.
\newblock {Parametrized measure models}.
\newblock {\em Bernoulli}, 24(3):1692 -- 1725, 2018.

\bibitem{B-R-2005}
D.~Beltita and T.~S. Ratiu.
\newblock {Symplectic leaves in real Banach Lie-Poisson spaces}.
\newblock {\em Geometric and Functional Analysis}, 15:753 -- 779, 2005.
\newblock {\href{https://doi.org/10.1007/s00039-005-0524-9}{DOI:
  10.1007/s00039-005-0524-9}}.

\bibitem{B-Z-2006}
I.~Bengtsson and K.~\.Zyczkowski.
\newblock {\em {Geometry of Quantum States: An Introduction to Quantum
  Entanglement}}.
\newblock Cambridge University Press, New York, 2006.
\newblock {\href{https://doi.org/10.1017/cbo9780511535048}{DOI:
  10.1017/cbo9780511535048}}.

\bibitem{Blackadar-2006}
B.~Blackadar.
\newblock {\em {Operator Algebras: Theory of $C^*$-algebras and von Neumann
  Algebras}}.
\newblock Springer-Verlag, Berlin, 2006.

\bibitem{B-R-1987-1}
O.~Bratteli and D.~W. Robinson.
\newblock {\em {Operator Algebras and Quantum Statistical Mechanics I}}.
\newblock Springer-Verlag, Berlin, second edition, 1987.
\newblock {\href{https://doi.org/10.1007/978-3-662-03444-6}{DOI:
  10.1007/978-3-662-03444-6}}.

\bibitem{C-I-M-M-2015}
J.~F. Cariñena, A.~Ibort, G.~Marmo, and G.~Morandi.
\newblock {\em {Geometry from dynamics, classical and quantum}}.
\newblock Springer, Dordrecht, 2015.
\newblock {\href{https://doi.org/10.1007/978-94-017-9220-2}{DOI:
  10.1007/978-94-017-9220-2}}.

\bibitem{Cencov-1982}
N.~N. Cencov.
\newblock {\em {Statistical Decision Rules and Optimal Inference}}.
\newblock American Mathematical Society, Providence, RI, 1982.
\newblock {\href{https://doi.org/10.1090/mmono/053}{DOI: 10.1090/mmono/053}}.

\bibitem{Choi-1974}
M.~Choi.
\newblock {A schwarz inequality for positive linear maps on
  $C^{\ast}$-algebras}.
\newblock {\em Illinois Journal of Mathematics}, 4(3):565--574, 1974.
\newblock {\href{https://doi.org/10.1215/ijm/1256051007}{DOI:
  10.1215/ijm/1256051007}}.

\bibitem{C-DC-I-M-2020-02}
F.~M. Ciaglia, F.~Di~Cosmo, A.~Ibort, and G.~Marmo.
\newblock {Schwinger's Picture of Quantum Mechanics}.
\newblock {\em International Journal of Geometric Methods in Modern Physics},
  17(04):2050054--14, 2020.
\newblock {\href{https://doi.org/10.1142/S0219887820500541}{DOI:
  10.1142/S0219887820500541}}.

\bibitem{C-I-M-2018}
F.~M. Ciaglia, A.~Ibort, and G.~Marmo.
\newblock {A gentle introduction to Schwinger's formulation of quantum
  mechanics: the groupoid picture}.
\newblock {\em Modern Physics Letters A}, 33(20):1850122--8, 2018.
\newblock {\href{https://doi.org/10.1142/s0217732318501225}{DOI:
  10.1142/s0217732318501225}}.

\bibitem{C-I-M-2019-02}
F.~M. Ciaglia, A.~Ibort, and G.~Marmo.
\newblock {Schwinger's Picture of Quantum Mechanics I: Groupoids}.
\newblock {\em International Journal of Geometric Methods in Modern Physics},
  16(08):1950119 (31), 2019.
\newblock {\href{https://doi.org/10.1142/S0219887819501196}{DOI:
  10.1142/S0219887819501196}}.

\bibitem{C-J-S-2020-02}
F.~M. Ciaglia, J.~Jost, and L.~Schwachh\"{o}fer.
\newblock {Differential geometric aspects of parametric estimation theory for
  states on finite-dimensional C*-algebras}.
\newblock {\em Entropy}, 22(11):1332, 2020.
\newblock {\href{https://doi.org/10.3390/e22111332}{DOI: 10.3390/e22111332}}.

\bibitem{C-J-S-2020}
F.~M. Ciaglia, J.~Jost, and L.~Schwachh\"{o}fer.
\newblock {From the Jordan product to Riemannian geometries on classical and
  quantum states}.
\newblock {\em Entropy}, 22(06):637--27, 2020.
\newblock {\href{https://doi.org/10.3390/e22060637}{DOI: 10.3390/e22060637}}.

\bibitem{C-J-S-2022}
F.~M. Ciaglia, J.~Jost, and L.~Schwachh\"{o}fer.
\newblock {What can Lie algebras tell us about Jordan algebras}.
\newblock {\em arXiv:2112.09781 [math.DG]}, 2021.
\newblock {\href{https://arxiv.org/abs/2112.09781}{arXiv}}.

\bibitem{C-M-P-1990}
R.~Cirelli, A.~Mania, and L.~Pizzocchero.
\newblock {Quantum mechanics as an infinite dimensional Hamiltonian system with
  uncertainty structure Part I}.
\newblock {\em Journal of Mathematical Physics}, 31(12):2891--2903, 1990.
\newblock {\href{https://doi.org/10.1063/1.528941}{DOI: 10.1063/1.528941}}.

\bibitem{DA-F-2021}
F.~D'Andrea and D.~Franco.
\newblock {On the pseudo-manifold of quantum states}.
\newblock {\em Differential Geometry and its Applications}, 78:101800, 2021.
\newblock {\href{https://doi.org/10.1016/j.difgeo.2021.101800}{DOI:
  10.1016/j.difgeo.2021.101800}}.

\bibitem{E-M-M-2010}
E.~Ercolessi, G.~Marmo, and G.~Morandi.
\newblock {From the equations of motion to the canonical commutation
  relations}.
\newblock {\em Rivista del Nuovo Cimento}, 33:401--590, 2010.
\newblock {\href{https://doi.org/10.1393/ncr/i2010-10057-x}{DOI:
  10.1393/ncr/i2010-10057-x}}.

\bibitem{F-F-M-P-2014}
P.~Facchi, L.~Ferro, G.~Marmo, and S.~Pascazio.
\newblock {Defining quantumness via the Jordan product}.
\newblock {\em Journal of Physics A: Mathematical and Theoretical}, 47(3),
  2014.
\newblock {\href{https://doi.org/10.1088/1751-8113/47/3/035301}{DOI:
  10.1088/1751-8113/47/3/035301}}.

\bibitem{F-F-I-M-2013}
F.~Falceto, L.~Ferro, .~Ibort, and G.~Marmo.
\newblock {Reduction of Lie-Jordan algebras and quantum states}.
\newblock {\em Journal of Physics A: Mathematical and Theoretical},
  46(1):015201--14, 2013.
\newblock {\href{https://doi.org/10.1088/1751-8113/46/1/015201}{DOI:
  10.1088/1751-8113/46/1/015201}}.

\bibitem{Fritz-2020}
T.~Fritz.
\newblock {A synthetic approach to Markov kernels, conditional independence and
  theorems on sufficient statistics}.
\newblock {\em Advances in Mathematics}, 370:107239, 2020.
\newblock {\href{https://doi.org/10.1016/j.aim.2020.107239}{DOI:
  10.1016/j.aim.2020.107239}}.

\bibitem{F-G-P-2021}
T.~Fritz, T.~Gonda, and P.~Perrone.
\newblock {De Finetti's Theorem in Categorical Probability}.
\newblock {\em arXiv math-PR: 2105.02639}, 2021.

\bibitem{F-G-P-R-2020}
T.~Fritz, T.~Gonda, P.~Perrone, and E.~F. Rischel.
\newblock {Representable Markov Categories and Comparison of Statistical
  Experiments in Categorical Probability}.
\newblock {\em arXiv math-st: 2010.07416}, 2020.

\bibitem{Fujiwara-1999}
A.~Fujiwara.
\newblock {Geometry of Quantum Information Systems}.
\newblock In O.~E. Barndorff-Nielsen and E.~B.~V. Jensen, editors, {\em
  {Geometry in Present Day Science}}, pages 35--48, 1999.
\newblock {\href{https://doi.org/10.1142/3958}{DOI: 10.1142/3958}}.

\bibitem{Garding-1947}
L.~G{\aa}rding.
\newblock {Note on Continuous Representations of Lie Groups}.
\newblock {\em Proceedings of the National Academy of Sciences of the United
  States of America}, 33(11):331--332, 1947.
\newblock {\href{https://doi.org/10.1073/pnas.33.11.331}{DOI:
  10.1073/pnas.33.11.331}}.

\bibitem{G-I-2001}
P.~Gibilisco and T.~Isola.
\newblock {A characterization of Wigner-Yanase skew information among
  statistically monotone metrics}.
\newblock {\em {Infinite Dimensional Analysis, Quantum Probability and Related
  Topics}}, 4(4):553--557, 2001.
\newblock {\href{https://doi.org/10.1142/s0219025701000644}{DOI:
  10.1142/s0219025701000644}}.

\bibitem{G-I-2003}
P.~Gibilisco and T.~Isola.
\newblock {Wigner-Yanase information on quantum state space: the geometric
  approach}.
\newblock {\em Journal of Mathematical Physics}, 44(9):3752--3762, 2003.
\newblock {\href{https://doi.org/10.1063/1.1598279}{DOI: 10.1063/1.1598279}}.

\bibitem{G-K-M-2006}
J.~Grabowski, M.~Ku{\'s}, and G.~Marmo.
\newblock {Symmetries, group actions, and entanglement}.
\newblock {\em Open Systems \& Information Dynamics}, 13(04):343 -- 362, 2006.

\bibitem{G-N-2020}
H.~Gzyl and F.~Nielsen.
\newblock {Geometry of the probability simplex and its connection to the
  maximum entropy method}.
\newblock {\em Journal of Applied Mathematics, Statistics and Informatics},
  16(01):25--35, 2020.
\newblock {\href{https://doi.org/10.2478/jamsi-2020-0003}{DOI:
  10.2478/jamsi-2020-0003}}.

\bibitem{G-R-2006}
H.~Gzyl and L.~Recht.
\newblock {A geometry on the space of probabilities I. The finite dimensional
  case}.
\newblock {\em Revista Matematica Iberoamericana}, 22(02):545--558, 2006.

\bibitem{G-R-2006-02}
H.~Gzyl and L.~Recht.
\newblock {A geometry on the space of probabilities II. Projective spaces and
  exponential families}.
\newblock {\em Revista Matematica Iberoamericana}, 22(03):833--849, 2006.

\bibitem{Hasegawa-2003}
H.~Hasegawa.
\newblock {Dual geometry of the Wigner-Yanase-Dyson information content}.
\newblock {\em Infinite Dimensional Analysis, Quantum Probability and Related
  Topics}, 6(3):413--430, 2003.
\newblock {\href{https://doi.org/10.1142/S021902570300133X}{DOI:
  10.1142/S021902570300133X}}.

\bibitem{Holevo-2001}
A.~S. Holevo.
\newblock {\em {Statistical Structure of Quantum Theory}}.
\newblock Springer-Verlag, Berlin, 2001.
\newblock {\href{https://doi.org/10.1007/3-540-44998-1}{DOI:
  10.1007/3-540-44998-1}}.

\bibitem{Holevo-2011}
A.~S. Holevo.
\newblock {\em {Probabilistic and Statistical Aspects of Quantum Theory}}.
\newblock Edizioni della Normale, 2011.
\newblock {\href{https://doi.org/10.1007/978-88-7642-378-9}{DOI:
  10.1007/978-88-7642-378-9}}.

\bibitem{Jencova-2006}
A.~Jen\v{c}ov\'{a}.
\newblock {A construction of a nonparametric quantum information manifold}.
\newblock {\em Journal of Functional Analysis}, 239(1):1--20, 2006.
\newblock {\href{https://doi.org/10.1016/j.jfa.2006.02.007}{DOI:
  10.1016/j.jfa.2006.02.007}}.

\bibitem{Jost-2017}
J.~Jost.
\newblock {\em {Riemannian Geometry and Geometric Analysis}}.
\newblock Springer, Berlin-Heidelberg, Germany, seventh edition, 2017.

\bibitem{Kibble-1979}
T.~W.~B. Kibble.
\newblock {Geometrization of Quantum Mechanics}.
\newblock {\em Communications in Mathematical Physics}, 65(2):189 -- 201, 1979.

\bibitem{Kirillov-1962}
A.~A. Kirillov.
\newblock {Unitary representations of nilpotent lie groups}.
\newblock {\em Russian Mathematical Surveys}, 17(4):53--104, 1962.
\newblock {\href{https://doi.org/10.1070/RM1962v017n04ABEH004118}{DOI:
  10.1070/RM1962v017n04ABEH004118}}.

\bibitem{Kirillov-1976}
A.~A. Kirillov.
\newblock {\em {Elements of the Theory of Representations}}.
\newblock Springer-Verlag Berlin Heidelberg, 1976.
\newblock {\href{https://doi.org/10.1007/978-3-642-66243-0}{DOI:
  10.1007/978-3-642-66243-0}}.

\bibitem{Kostant-1970}
B.~Kostant.
\newblock {Quantization and unitary representations}.
\newblock In {\em {Lectures in Modern Analysis and Applications III}}, volume
  170 of {\em Lecture Notes in Mathematics}, pages 87--208. Springer, Berlin,
  Heidelberg, 1970.
\newblock {\href{https://doi.org/10.1007/BFb0079068}{DOI: 10.1007/BFb0079068}}.

\bibitem{Landsman-1998}
N.~P. Landsman.
\newblock {\em {Mathematical Topics Between Classical and Quantum Mechanics}}.
\newblock Springer, New York, NY, 1998.
\newblock {\href{https://doi.org/10.1007/978-1-4612-1680-3}{DOI:
  10.1007/978-1-4612-1680-3}}.

\bibitem{Landsman-2017}
N.~P. Landsman.
\newblock {\em {Foundations of Quantum Theory. From Classical Concepts to
  Operator Algebras}}.
\newblock Springer International Publishing, Cham, Switzerland, 2017.
\newblock {\href{https://doi.org/10.1007/978-3-319-51777-3}{DOI:
  10.1007/978-3-319-51777-3}}.

\bibitem{Lang-1999}
S.~Lang.
\newblock {\em {Fundamentals of Differential Geometry}}.
\newblock Springer-Verlag, Berlin, 1999.

\bibitem{L-Y-L-W-2020}
J.~Liu, H.~Yuan, X.-M. Lu, and X.~Wang.
\newblock {Quantum Fisher information matrix and multiparameter estimation}.
\newblock {\em Journal of Physics A: Mathematical and Theoretical},
  53(2):023001--69, 2020.
\newblock {\href{https://doi.org/10.1088/1751-8121/ab5d4d}{DOI:
  10.1088/1751-8121/ab5d4d}}.

\bibitem{M-N-1987}
G.~Maltese and G.~Niestegge.
\newblock {A linear Radon-Nikodym type theorem for $C^{*}$-algebraswith
  applications to measure theory}.
\newblock {\em Annali della Scuola Normale Superiore di Pisa, Classe di Scienze
  4}, 14(2):345--354, 1987.

\bibitem{M-M-V-V-2017}
V.~I. Man'ko, G.~Marmo, F.~Ventriglia, and P.~Vitale.
\newblock {Metric on the space of quantum states from relative entropy.
  Tomographic reconstruction}.
\newblock {\em Journal of Physics A: Mathematical and Theorerical},
  50(33):335302, 2017.
\newblock {\href{https://doi.org/10.1088/1751-8121/aa7d7d}{DOI:
  10.1088/1751-8121/aa7d7d}}.

\bibitem{Michor-1980}
Peter~W Michor.
\newblock {\em {Manifolds of differentiable mappings}}.
\newblock Shiva Publishing Limited, 1980.
\newblock {\href{https://doi.org/10.1007/978-3-642-11102-0_5}{DOI:
  10.1007/978-3-642-11102-0\textunderscore 5}}.

\bibitem{N-V-W-1975}
J.~Naudts, A.~Verbeure, and R.~Weder.
\newblock {Linear Response Theory and the KMS Condition}.
\newblock {\em Communications in Mathematical Physics}, 44:87--99, 1975.
\newblock {\href{https://doi.org/10.1007/BF01609060}{DOI: 10.1007/BF01609060}}.

\bibitem{N-C-2011}
M.~A. Nielsen and I.~L. Chuang.
\newblock {\em {Quantum Computation and Quantum Information}}.
\newblock Cambridge University Press, New York, NY, 2011.

\bibitem{Niestegge-1983}
G.~Niestegge.
\newblock {Absolute continuity for linear forms on $B^{*}$-algebras and a
  Radon-Nikodym type theorem (quadratic version)}.
\newblock {\em Rendiconti del Circolo Matematico di Palermo}, 32(2):358--376,
  1983.

\bibitem{Paris-2009}
M.~G.~A. Paris.
\newblock {Quantum Estimation for Quantum Technology}.
\newblock {\em International Journal of Quantum Information}, 7(1):125--137,
  2009.
\newblock {\href{https://doi.org/10.1142/S0219749909004839}{DOI:
  10.1142/S0219749909004839}}.

\bibitem{Parzygnat-2020}
A.~J. Parzygnat.
\newblock {Inverses, disintegrations, and Bayesian inversion in quantum Markov
  categories}.
\newblock {\em arXiv quant-ph: 2001.08375}, 2020.
\newblock {\href{https://doi.org/10.48550/arXiv.2001.08375}{DOI:
  10.48550/arXiv.2001.08375}}.

\bibitem{Perelomov-1986}
A.~Perelomov.
\newblock {\em Generalized Coherent States and Their Applications}.
\newblock Springer-Verlag Berlin Heidelberg, 1986.

\bibitem{Petz-1994}
D.~Petz.
\newblock Geometry of canonical correlation on the state space of a quantum
  system.
\newblock {\em Journal of Mathematical Physics}, 35(2):780--795, 1994.
\newblock {\href{https://doi.org/10.1063/1.530611}{DOI: 10.1063/1.530611}}.

\bibitem{Petz-1996}
D.~Petz.
\newblock {Monotone metrics on matrix spaces}.
\newblock {\em Linear Algebra and its Applications}, 244:81--96, 1996.
\newblock {\href{https://doi.org/10.1016/0024-3795(94)00211-8}{DOI:
  10.1016/0024-3795(94)00211-8}}.

\bibitem{Petz-2007}
D.~Petz.
\newblock {\em {Quantum information theory and quantum statistics}}.
\newblock Springer, Berlin, 2007.

\bibitem{P-S-1995}
G.~Pistone and C.~Sempi.
\newblock {An infinite-dimensional geometric structure on the space of all the
  probability measures equivalent to a given one}.
\newblock {\em The Annals of Statistics}, 23(5):1543 -- 1561, 1995.

\bibitem{Sakai-1997}
S.~Sakai.
\newblock {\em {$C^{*}$-algebras and $W^{*}$-algebras}}.
\newblock Springer-Verlag, Berlin, 1997.
\newblock {\href{https://doi.org/10.1007/978-3-642-61993-9}{DOI:
  10.1007/978-3-642-61993-9}}.

\bibitem{Souriau-1970}
J.-M. Souriau.
\newblock {\em {Structure des systèmes dynamiques}}.
\newblock Dunod, Paris, 1970.
\newblock
  {\href{http://www.jmsouriau.com/structure_des_systemes_dynamiques.htm}{Web
  source}}.

\bibitem{Suzuki-2019}
J.~Suzuki.
\newblock {Information Geometrical Characterization of Quantum Statistical
  Models in Quantum Estimation Theory}.
\newblock {\em Entropy}, 21(7):703, 2019.
\newblock {\href{https://doi.org/10.3390/e21070703}{DOI: 10.3390/e21070703}}.

\bibitem{Takesaki-2002}
M.~Takesaki.
\newblock {\em {Theory of Operator Algebra I}}.
\newblock Springer-Verlag, Berlin, 2002.

\end{thebibliography}}
\end{document}